\documentclass[fleqn,usenatbib]{mnras}

\usepackage{newtxtext,newtxmath}

\usepackage[T1]{fontenc}

\DeclareRobustCommand{\VAN}[3]{#2}
\let\VANthebibliography\thebibliography
\def\thebibliography{\DeclareRobustCommand{\VAN}[3]{##3}\VANthebibliography}

\usepackage{graphicx}	
\usepackage{amsmath}	

\def\dif{\mathrm{d}}

\def\qq{\boldsymbol{q}}

\def\xx{\boldsymbol{x}}

\def\ihMpc{h \, \mathrm{Mpc}^{-1}}
\def\hMpc{h^{-1} \, \mathrm{Mpc}}
\def\narya{\texttt{Narya}}
\def\nenya{\texttt{Nenya}}
\def\vilya{\texttt{Vilya}}
\def\theone{\texttt{TheOne}}

\newcommand{\ensavg}[1]{\left\langle #1 \right\rangle}

\title[Priors on Lagrangian bias]{Priors on Lagrangian bias parameters from galaxy formation modelling}

\author[Zennaro et al.]{
Matteo Zennaro,$^{1}$\thanks{E-mail:matteo\_zennaro001@ehu.eus}
Raul E. Angulo,$^{1,2}$
Sergio Contreras,$^{1}$
Marcos Pellejero-Ib\'a\~nez,$^{1}$
Francisco Maion$^{1,3}$
\\
$^{1}$Donostia International Physics Center (DIPC), Paseo Manuel de Lardizabal, 4, 20018, Donostia-San Sebasti\'an, Guipuzkoa, Spain.\\
$^{2}$IKERBASQUE, Basque Foundation for Science, 48013, Bilbao, Spain.\\
$^{3}$Departamento de F\'isica Matem\'atica, Instituto de F\'isica, Universidade de S\~ao Paulo, Rua do Mat\~ao 1371, CEP 05508-090, S\~ao Paulo, Brazil.
}

\date{Accepted XXX. Received YYY; in original form ZZZ}

\pubyear{2020}
\begin{document}
\label{firstpage}
\pagerange{\pageref{firstpage}--\pageref{lastpage}}
\maketitle

\begin{abstract}
     We study the relations among the parameters of the hybrid Lagrangian bias expansion model, fitting biased auto and cross power spectra up to $k_{\rm max} = 0.7 \ihMpc$. We consider $\sim 8000$ halo and galaxy samples, with different halo masses, redshifts, galaxy number densities, and varying the parameters of the galaxy formation model. Galaxy samples are obtained through state-of-the-art extended subhalo abundance matching techniques and include both stellar-mass and star-formation-rate selected galaxies. All of these synthetic galaxies samples are publicly available at \url{https://bacco.dipc.org/galpk.html}. We find that the hybrid Lagrangian bias model provides accurate fits to all of our halo and galaxy samples. The coevolution relations between galaxy bias parameters, although roughly compatible with those obtained for haloes, show systematic shifts and larger scatter. We explore possible sources of this difference in terms of dependence on halo occupation and assembly bias of each sample. The bias parameter relations displayed in this work can be used as a prior for future Bayesian analyses employing the hybrid Lagrangian bias expansion model.
\end{abstract}

\begin{keywords}
cosmology: theory -- large-scale structure of the Universe -- methods: statystical -- methods: computational
\end{keywords}



\section{Introduction}
The coming years will be marked by the first light of many astronomical instruments, which are expected to produce the most precise observations of the clustering of galaxies to date. These observational campaigns will include spectroscopic and photometric data, covering unprecedented cosmological volumes, unlocking, at the same time, information at the smallest scales ever observed in cosmology.

Along with this incredible advance on the observational side, the modelling of the clustering of galaxies from a theoretical standpoint is also being pushed to its limits. As a matter of fact, even assuming to know the nonlinear growth of dark matter, a worrisome unknown is represented by galaxy formation.

However, one can try to directly model galaxy formation processes in simulations, solving hydrodynamical equations that represent the evolution of gas and stars, and adding a number of sub-grid processes to model the main drivers of galaxy evolution, dynamics and quenching \citep[see][]{VogelsbergerEtal2014,DuboisEtal2014,SchayeEtal2015,DaveEtal2019}. This process is of course quite computationally expensive, and is therefore used for relatively small-size simulations, run assuming few, very sensible, cosmologies. Semi-analytic models \citep[for example,][]{HenriquesEtal2015,StevensEtal2016,LaceyEtal2016,CrotonEtal2016,LagosEtal2018} can partially alleviate the overall computational cost by adding gas and stars in post-production, following a number of empirical and semi-empirical relations. This could still be quite expensive if the goal is efficiently creating fully nonlinear galaxy catalogues that could be directly compared to observations. One of the most promising methods in this sense is the so called sub-halo abundance matching \citep{ConroyEtal2006,ReddickEtal2013,Chaves-MonteroEtal2016,LehmannEtal2017,DragomirEtal2018}, which exploits the substructures identified in gravity-only simulations, matching their abundance to an observed stellar-mass or luminosity functions. Recent improvements to this technique, such as the SHAMe model proposed by \cite{ContrerasAnguloZennaro2021b}, have proven particularly reliable in reproducing hydrodynamical results at a fraction of the computational cost.

A completely different approach is to model galaxy clustering in a way that is agnostic of the galaxy formation model \citep[see the review on bias by][]{DesjacquesJeongSchmidt2016}. In this respect, the only assumption is that the galaxy field is a function of the underlying matter density and velocity distribution. The accuracy of this model will depend on the quality of the description of the underlying matter field, and the robustness of the chosen biasing function. There are many possible choices to describe the biasing scheme: it can be described in the final Eulerian coordinates or in the Lagrangian coordinates corresponding to the initial state of the matter field; it can include expansions of the density field in powers; it can account for the effect of the tidal field and other nonlocal quantities; it can account for derivatives of the density field. While directly modelling galaxy formation can put constraints on the physical processes at play, it requires us to develop a model of those physical processes, i.e. a different model for each type of biased tracers. The more agnostic approach has the advantage of being very flexible, so the same model can describe biased tracers with extremely different properties.

In particular, approaches in the context of perturbation theory and effective field theory allow us to describe the clustering of matter and galaxies down to scales around $k_{\rm max} \sim 0.2 \ihMpc$ \citep{BaumannEtal2012,BaldaufSchaanZaldarriaga2016,VlahCastorinaWhite2016,IvanovSimonovicZaldarriaga2020b,DamicoEtal2020,ColasEtal2020,NishimichiEtal2020,ChenVlahWhite2020}, which can be pushed to $k_{\rm max} \sim 0.3 \ihMpc$ when combining 2- and 3-points statistics while including 1-loop corrections \citep[see, e.g.,][]{EggemeierEtal2021,PezzottaEtal2021}. Cosmological simulations, on the other hand, can capture the gravitational growth of structure in the universe down to very small scales; however, they can quickly become prohibitively expensive, especially when we want to resolve very small scales while keeping the probed volume large enough to neglect sample variance. Finally, some approaches have been proposed to create hybrid descriptions of the nonlinear galaxy density fields that rely on perturbation theory to describe large scales (avoiding finite-volume problems present in simulations), but describe smaller, nonlinear scales using numerical $N$-body solutions \citep{ModiChenWhite2020}. One preliminary but very promising application of this method was presented in \cite{HadzhiyskaEtal2021}, where the authors successfully employed it to describe DES Y1 lensing shear and projected galaxy clustering, including nonlinear scales until $k_{\rm max} = 0.5 \, \ihMpc$, and shrinking by 35\% the uncertainty on $\Omega_{\rm m}$ with respect to previous analyses.

In this more agnostic approach, one problem is the increase of free parameters in the model, which could translate into lower constraining power. However, bias parameters are not fully unrelated. In particular, many works studying samples of mass-selected haloes found that higher order bias parameters can be related to the linear bias. These relations can be purely empirical \citep[][]{TinkerEtal2010, LazeyrasEtal2016, LazeyrasSchmidt2018}, or rely on some model of structure collapse \citep[][]{ShethChanScoccimarro2013}. We note that all these relations are obtained considering the large-scale limit of these bias parameters.

In addition, a recent work by \cite{BarreiraLazeyrasSchmidt2021} has shown, employing a forward modelling of galaxies in the context of a Eulerian bias expansion, that these relations also hold (to good approximation) in the case of galaxies selected from the TNG hydrodynamical simulation. The authors notice slight deviations from the results obtained from haloes, that they interpret in the context of the halo model and galaxy assembly bias.

In this work, we study the relations between bias parameters considering a Lagrangian bias expansion up to second order. In particular, to model the  auto and cross power spectra of the biased tracers we adopt the hybrid approach described in \cite{ZennaroEtal2021}, thus pushing the determination of the bias parameters to scales as small as $k_{\rm max} \sim 0.7 \ihMpc$. We consider as biased tracers haloes and galaxies from $N$-body simulations in four cosmologies. For galaxies, we adopt an extended subhalo abundance matching technique that allows us to vary the galaxy formation parameters exploring a wide variety of models, both selecting galaxies by stellar mass and by star-formation rate.

Furthermore, we create ad-hoc galaxy mock catalogues using a halo occupation distribution (HOD) technique to separately study the effect of different halo occupations and of galaxy assembly bias (GAB) on the bias relations.

This paper is organized as follows: in section \ref{sec::sims-mocks} we present our simulations and galaxy mocks; in section \ref{sec::model} we review the hybrid Lagrangian bias expansion model; in section \ref{sec::fits} we illustrate the details of our fitting procedures and address its accuracy and limitations; we present our results for haloes, and stellar mass selected and star-formation rate selected galaxies in section \ref{sec::results}, and we discuss these results and their interpretation in section \ref{sec::discussion}; finally, in section \ref{sec::conclu} we present our conclusions.

\section{Simulations and galaxy mocks}\label{sec::sims-mocks}
In the present section, we will introduce the dark matter simulations employed in this work. Moreover, we will describe the procedure we followed to obtain galaxy samples from these simulations, as well as the characteristics of the galaxy mocks.

\subsection{Dark-matter simulations}
In this work we use the eight BACCO simulations presented for the first time in \cite{AnguloEtal2021}. These are four pairs of simulations, each pair assuming one of the main BACCO cosmologies (\narya, \nenya, \vilya, and \theone, see Tab. \ref{tab:cosmologies}). The two realisations composing each pair are characterised by same fixed-amplitude initial fields with opposite phases, employing the ``Fixed \& Paired'' technique \citep{AnguloPontzen2016}, that allows us to suppress cosmic variance by at least two orders of magnitudes on scales $k < 0.1 \ihMpc$.

\begin{table}
    \centering
    \begin{tabular}{cc|cccccccc} 
       \hline
       Cosmology &  $\Omega_{\rm cdm}$ & $\Omega_{\rm b}$ & $h$ & $n_{\rm s}$\\
       \hline
      \nenya  & 0.265 & 0.050 & 0.60 & 1.01 \\
      \narya  & 0.310 & 0.050 & 0.70 & 1.01 \\
      \vilya  & 0.210 & 0.060 & 0.65 & 0.92 \\
      \theone & 0.259 & 0.048 & 0.68 & 0.96 \\
       \hline
       \end{tabular}
      \caption{Cosmological parameters of the four main cosmologies of the BACCO project. All of them assume a flat geometry, no massive neutrinos ($M_{\nu}=0$ eV), a dark energy equation of state with $w_0=-1$ and $w_a=0$, an amplitude of cold matter fluctuations $\sigma_8=0.9$, and optical depth at recombination $\tau=0.0952$.}
    \label{tab:cosmologies}
\end{table}

Each simulation follows the evolution of $4320^3$ cold matter particles in a comoving cubical box of side $L_{\rm box} = 1440 \hMpc$. The particle mass is roughly $3 \times 10^9 h^{-1} M_\odot$. The initial positions and velocities are set at $z=49$ with second-order Lagrangian Perturbation Theory.

Both the set up of the initial conditions and the gravitational evolution of these simulations are performed using the code \texttt{L-Gadget3} \citep{AnguloEtal2012,AnguloEtal2021}, a lean version of \texttt{Gadget} \citep{Springel2005}. The Plummer-equivalent softening length and the other precision parameters of the code have been set in accordance to \cite{AnguloEtal2021}, in order to achieve a convergence of the matter power spectrum at 2\% level at $k\sim 10 \ihMpc$.

A key aspect of \texttt{L-Gadget3} is that it features an improved version of the substructure finding algorithm \texttt{SUBFIND} \citep{DavisEtal1985}. In particular, \texttt{L-Gadget3} is able to find haloes and subhaloes on the fly, storing as well a number of (sub)halo properties that are nonlocal in time (such as the peak mass or circular velocity ever attained by each substructure). This is particularly useful for efficiently building mock catalogues of galaxies.

Besides our main simulations, we also use a set of smaller-volume simulations. We will employ them in Section \ref{sec::discussion} to explore some possible origins of the difference between galaxy and halo coevolution relations, with galaxy samples of more manageable size. These simulations have the same cosmologies and mass resolutions as their larger counterparts. In this case, the side of each comoving box is $512 \hMpc$, and each samples the dark matter distribution with $1536^3$ particles.

\subsection{Galaxy catalogues}
We create our galaxy catalogues applying the SubHalo Abundance Matching extended algorithm (SHAMe), presented in \cite{ContrerasAnguloZennaro2021b}. We refer the reader to the SHAMe presentation paper for a more detailed discussion of its features, while we limit ourselves to underline the parts relevant to the present work. In particular, within the SHAMe context, we are able to create two types of galaxy catalogues.

By matching the abundance of haloes selected according to their peak circular velocity $v_{\rm peak}$ to the stellar-mass ($M_*$) function of the TNG-300 hydrodynamical simulation \citep{NelsonEtal2018,SpringelEtal2018,MarinacciEtal2018,PillepichEtal2018,NaimanEtal2018} we obtain \textit{stellar-mass (SM) selected} galaxy samples. We remark that, by construction, the galaxy clustering obtained with the SHAMe model does not depend on the choice of stellar mass function, as long as the subhalo rank ordering is preserved. In this case, we have control on three aspects of the model (corresponding to three free parameters):
\begin{itemize}
    \item the scatter in the $v_{\rm peak}-M_*$ relation, $\sigma_{M_*}$;
    \item the survival time of orphan subhaloes $t_{\rm merger}$ before we consider them completely merged with their host halo;
    \item the fraction of the peak mass $f_{\rm s}$ of a subhalo at which we consider it dynamically disrupted.
\end{itemize}
While the former parameter mostly shuffles haloes among bias/mass bins, the latter two affect mostly the satellite fraction of the sample.

The second kind of galaxy sample we can obtain with SHAMe is \textit{star-formation rate (SFR) selected} galaxies. To this purpose, we match the peak mass $m_{\rm peak}$ of each subhaloes to the star-formation rate  predicted for each structure according to the empirical prescriptions of the SHAMe model. In this case there are five free parameters:
\begin{itemize}
    \item the slopes $\beta$ and $\gamma$ of the broken power law describing the star-formation efficiency;
    \item the mass of peak star formation efficiency, $M_1$;
    \item the timescale of the star-formation quenching after a given halo has been accreted by a larger halo, governed by the parameters $\tau_0$ and $\tau_{\rm s}$.
\end{itemize}

In both cases, after building the rank-ordered galaxy catalogue, we apply a cut to select a given number density. This means that the method does not heavily depend neither on the specific SM function nor on the SFR model adopted, as long as the resulting rank-ordering is not changed. We consider galaxy samples with four, very different, number density cuts, namely $\bar{n} = \{0.01, 0.00316, 0.001, 0.0003\} h^3 \, \mathrm{Mpc}^{-3}$. The different number densities and selection criteria allow us to explore samples whose characteristics realistically span the variety expected from current and upcoming galaxy surveys. While the densest SM-selected sample could represent a SPHEREX-like survey at low redshift, the intermediate SM-selected sample is more similar to the CMASS sample of BOSS, and the sparsest SFR-selected samples reflect a Euclid-like survey \citep{DoreEtal2014,Rodriguez-TorresEtal2016,EuclidValidation2020}.

One interesting feature of SHAM methods is that galaxies obtained with these techniques present a certain amount of galaxy assembly bias, that is the dependence of galaxy clustering on properties other than the host halo mass. In \cite{Chaves-MonteroEtal2016}, the authors showed that SHAM galaxies could reproduce a significative fraction of the total assembly bias signal present in the EAGLE hydrodynamical simulations (specifically they find a 20\% assembly bias signal in the EAGLE galaxy sample and the same signal appears at 15\% level in the SHAM sample). In principle, the extended SHAM version adopted in this work allows for mimicking any desired amount of galaxy assembly bias signal, fudging two extra free parameters \citep{ContrerasAnguloZennaro2021a}. However, since we are not trying to reproduce any specific observation, we will not take advantage of this possibility, and we will just note that our samples do present a galaxy assembly bias contribution, which is different in each case depending on the galaxy formation parameters.

\begin{figure*}
    \includegraphics[width=0.48\textwidth]{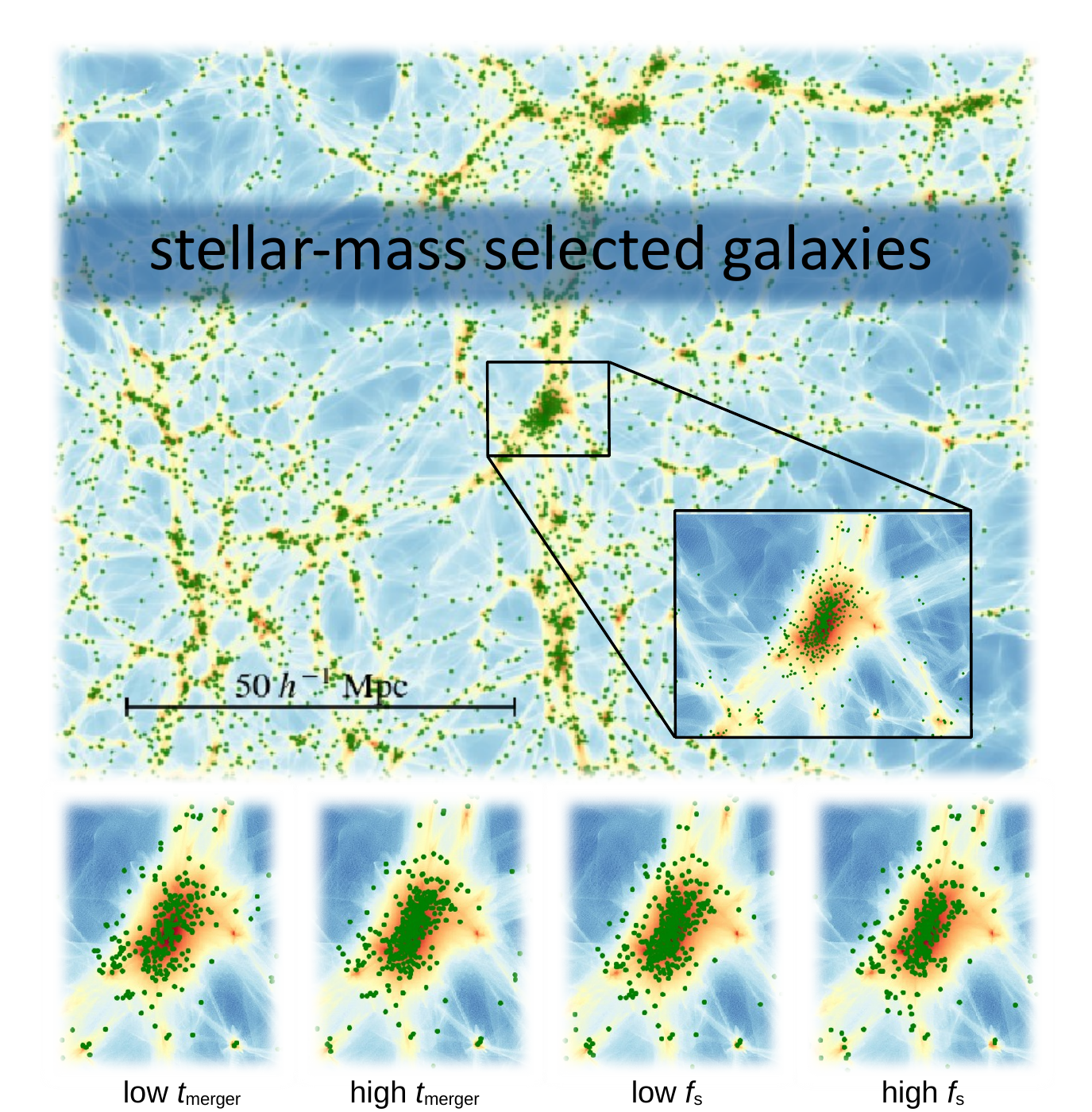}
    \includegraphics[width=0.48\textwidth]{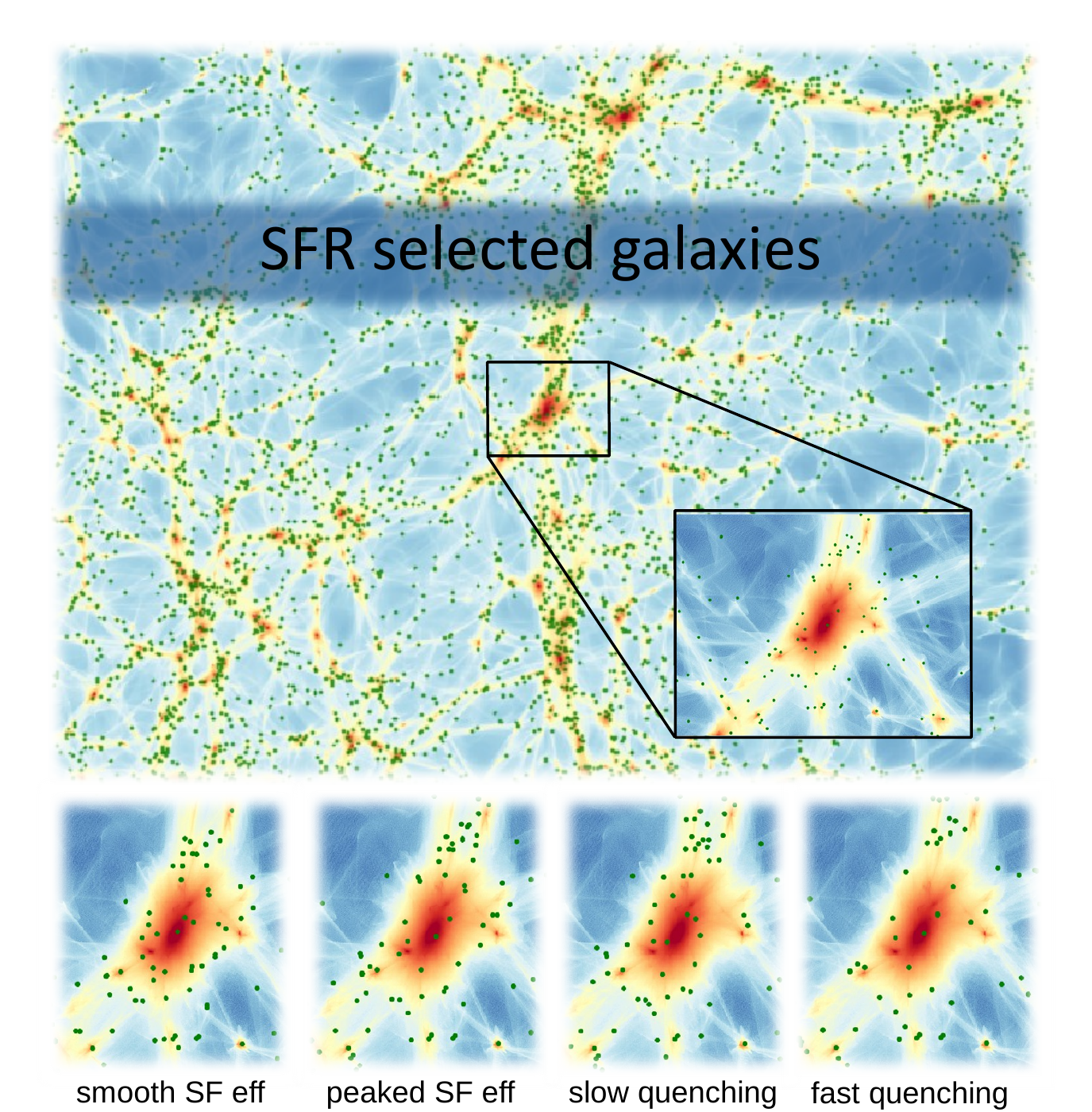}
    \caption{A 128 $\hMpc$ wide view of the distribution of dark matter and galaxies around a massive halo at $z=0$. In all cases, a galaxy sample with number density $n=0.01 h^3 \mathrm{Mpc}^{-3}$ has been selected. The dark matter field is coloured from blue to red for increasing density. Galaxy positions are marked with green points. On the left, galaxies have been selected according to their stellar mass. In the lower panels we show how the central halo is populated differently when we change $t_{\rm merger}$ (with low values, orphan subhaloes are quickly suppressed, see two leftmost sub-panels), or when we modify $f_s$ (with low $f_s$ subhaloes are never dynamically disrupted, while high values cause the satellite fraction to drastically decrease, see rightmost two sub-panels). On the right, the galaxy SFR is the selection criterion. In this case, in the two leftmost lower panels we show the effect of varying $\beta$ and $\gamma$ (when both approach 0, the start formation efficiency is smooth, while when they are both large it is very peaked at the host mass $M_1$), while in the two rightmost panels we show the effect of $\tau_0$ and $\tau_{\rm S}$ (which regulate the rapidness of the SF quenching in satellite galaxies).}
    \label{fig::galaxies}
\end{figure*}

In Fig. \ref{fig::galaxies} we present a visualisation of two of our galaxy samples. On the left we show how SM-selected galaxies populate a $128 \hMpc$ wide region of a simulation at $z=0$. In particular it is possible to appreciate how galaxies with large stellar mass are more clustered than their underlying dark matter distribution and tend to preferentially be located in haloes and filaments. In the smaller panels below we also show a zoom-in on the largest halo of the simulation, varying some free parameters of the SHAMe model. In particular, we show how by increasing $f_{\rm s}$ we decrease the satellite fraction depopulating preferentially the outer parts of the halo, while a shorter merger time $t_{\rm merger}$ decreases the satellite fraction depleting also the innermost region of the halo.

In the right part of Fig. \ref{fig::galaxies} we show the same visualization for star forming galaxies. In this case galaxies are mostly found outside of dark matter haloes, either in lower density filaments or even in isolation \citep[see, for example,][]{OrsiAngulo2018}. For this reason changing the star formation efficiency and even the dynamical quenching time affects less the satellite fraction of host haloes, as shown in the lower panels.

For each of the four cosmologies considered we created 125 SM-selected and 125 SFR-selected galaxy samples. Each galaxy mock assumes a set of SHAMe parameters drawn from a 3-dimensional latin hypercube for the SM-selected sample, and from a 5-dimensional latin hypercube for the SFR-selected sample. The parameter space of the two latin hypercubes is reported in Table \ref{tab:LH}. It corresponds to the choice presented in \cite{ContrerasAnguloZennaro2021b}, where these values were specifically designed to span a significantly larger parameter space than what is currently allowed by hydrodynamic simulations. We repeat this procedure at redshifts $z=0$ and $z=1$. From each galaxy mock we extract four subsamples with our fiducial number densities.

Finally, we measure the galaxy-galaxy auto power spectrum and the galaxy-matter cross power spectrum assigning each of these distributions to a grid with $N_{\rm grid} = 1080^3$ with a Cloud-in-Cell mass assignment scheme. To represent each field, we use two interlaced grids to reduce the effect of aliasing \citep{SefusattiEtal2016}.

With four cosmologies, two redshifts, 250 galaxy formation models and four number densities we have access to 8000 auto and cross power spectra that capture the vast diversity of possible realistic galaxy samples.

\begin{table}
    \centering
    \begin{tabular}{l c r}
        \hline
        \multicolumn{2}{c}{SM-selected samples}\\
        \hline
        $\sigma_{M_*}$ & $\in$ & [0, 0.4] \\
        $t_{\rm merger}$ & $\in$ & [0.01, 3] \\
        $f_{\rm s}$ & $\in$ & [0, 0.4] \\
        \hline
        \multicolumn{2}{c}{SFR-selected samples}\\
        \hline
        $\beta$  & $\in$& [0.1, 12] \\
        $\gamma$ & $\in$ & [0.1, 12] \\
        $\log_{10}M_1$ & $\in$ & [11.5, 13.5] \\
        $\tau_{0}$ & $\in$ & [0, 16] \\
        $\tau_{\rm s}$ & $\in$ & [-1, 0] \\
        \hline
    \end{tabular}
    \caption{The hypervolume covered by the two latin hypercubes used to select 125 random sets of parameters for SM-selected galaxies, and 125 random sets of parameters for star forming galaxies.}
    \label{tab:LH}
\end{table}

\section{Biasing model}\label{sec::model}
To describe our galaxy auto and cross power spectra we adopt a second-order Lagrangian bias expansion \citep{Matsubara2008}. In this context, the galaxy overdensity field at the Eulerian coordinates $\xx$ is expressed as a weighted advection of all contributions from initial (Lagrangian) coordinates $\qq$ that end up in $\xx$, that is
\begin{equation}
    1 + \delta_{\rm g} (\xx) = \int \dif^3 \qq w(\qq) \delta_{\rm D}(\xx - \qq - \boldsymbol{\psi}),
\end{equation}
where we have introduced the weighting function $w(\qq)$. This function can be computed at second order as the superposition of five fields (homogeneous, linear density, squared density, tidal and laplacian of the density), each weighted by its corresponding bias parameter,
\begin{equation}
    \begin{split}
    w(\qq) &= 1 + b_1^{\rm L} \delta(\qq) + b_2^{\rm L} \left[\delta^2(\qq) - \ensavg{\delta^2}\right]\\
    &+ b_{s^2}^{\rm L} \left[s^2(\qq) - \ensavg{s^2}\right] + b_{\nabla^2}^{\rm L} \nabla^2\delta(\qq),\\
    \end{split}
\end{equation}
where $s^2(\qq)$ is the traceless contracted tidal field, and $b_1^{\rm L}, b_2^{\rm L}, b_{s^2}^{\rm L},$ and $b_{\nabla^2\delta}^{\rm L}$ are the Lagrangian bias parameters, assumed to be scale independent in Lagrangian space.

After advecting to Eulerian coordinates and Fourier transforming the fields, the corresponding galaxy auto power spectrum reads
\begin{equation}
    P_{\rm gg}(k) = \sum_{i,j \in \{1, \delta, \delta^2, s^2, \nabla^2\delta\}} b_i^{\rm L} b_j^{\rm L} P_{ij}(k),
    \label{eq:auto}
\end{equation}
where $P_{ij}$ is the cross spectrum of the different fields. Note that by definition the bias parameter associated with the homogeneous field is 1, $b_{i=1}=1$, the parameter associated with the linear density field is $b_{i=\delta} = b_1$, and the one associated with the squared density field is $b_{i=\delta^2}=b_2$. Moreover, the galaxy-matter cross power spectrum will be simply given by
\begin{equation}
    P_{\rm gm}(k) = \sum_{i \in \{1, \delta, \delta^2, s^2, \nabla^2\delta\}} b_i^{\rm L} P_{1i}(k).
    \label{eq:cross}
\end{equation}

The model depends on the 15 cross-field Lagrangian terms advected to Eulerian coordinates $P_{ij}$. One approach to obtain them is to predict these terms using Lagrangian perturbation theory \citep{McEwenEtal2016,FangEtal2017,ChenVlahWhite2020,ZennaroEtal2021}. Another possibility is to measure them directly in simulations, and possibly combine the measurements with the perturbative solutions to suppress the noise that might be present on large scales. The latter approach was proposed initially by \citep{ModiChenWhite2020}, and advanced by \cite{KokronEtal2021} and \cite{ZennaroEtal2021}, as both works presented an emulator for this basis of 15 spectra.

In this work we measure the 15 Lagrangian fields directly in our 4 paired simulations and we match them on large scales with their corresponding perturbative solution. Lagrangian fields are more affected by exclusion effects than their Eulerian counterparts, since the Lagrangian volume occupied by collapsed objects is significantly larger than its corresponding Eulerian volume. We alleviate this problem by smoothing our Lagrangian fields. Unless otherwise stated, we always assume a smoothing scale $k_{\rm d} = 0.75 \ihMpc$, applied to the linear power spectrum employed to create the Lagrangian fields. We remark that the chosen smoothing scales also sets a hard limit to the smallest scales that can be included in our analysis, since we need to ensure that $k_{\rm max} < k_{\rm d}$ at all times.

\section{Fitting bias parameters}\label{sec::fits}

To infer the values assumed by the bias parameters in the different galaxy populations, we use as our data the measurements of the auto and cross galaxy power spectra $\tilde{P}(k) = \{P_{\rm gg}(k), P_{\rm gm}(k)\}$.
Our model $\tilde{P}_{\rm model}(k) = \{P_{\rm gg, model}(k), P_{\rm gm, model}(k)\}$ is given by a modified version of Eq.s (\ref{eq:auto}) and (\ref{eq:cross}), where we add a free parameter accounting for the contribution of shot noise,
\begin{equation}
    \begin{split}
        &P_{\rm gg, model}(k) = \sum_{i,j \in \{1, \delta, \delta^2, s^2, \nabla^2\delta\}} b_i^{\rm L} b_j^{\rm L} P_{ij}(k) + \dfrac{A_{\rm sn}}{\bar{n}},\\
        &P_{\rm gm, model}(k) = \sum_{i \in \{1, \delta, \delta^2, s^2, \nabla^2\delta\}} b_i^{\rm L} P_{1i}(k).
    \end{split}
\end{equation}
This means that in the Eulerian auto power spectrum we assume that the shot noise contribution comes from a Poissonian distribution, with amplitude given by $A_{\rm sn}$.
The 5 free parameter therefore are
\begin{equation}
    \boldsymbol{\vartheta} = \left\{b^{\rm L}_1, b^{\rm L}_2, b^{\rm L}_{s^2}, b^{\rm L}_{\nabla^2\delta}, A_{\rm sn}\right\}.
\end{equation}
We vary these parameters in the hypervolume defined by
\begin{displaymath}
    \begin{split}
        &b_1^{\rm L} \in [-5, 20], \qquad & b_2^{\rm L} \in [-5, 10], \\
        &b_{s^2}^{\rm L} \in [-10, 20], \qquad & b_{\nabla^2\delta}^{\rm L} \in [-20, 30], \\
        &A_{\rm sn} \in [0, 2].
    \end{split}
\end{displaymath}
We assume that the likelihood of observing a particular set of power spectra given the model parameters is given by a multivariate normal distribution with
\begin{equation}
    \ln p[\tilde{P}(k) | \boldsymbol{\vartheta}] = - \dfrac{1}{2} \sum_{i} \left[\dfrac{\tilde{P}(k_i) - \tilde{P}_{\rm model}(k_i)}{\sigma_{k_i}^2}\right]^2 + c,
\end{equation}
where we treat the data covariance as diagonal. In particular, in the present work we do not consider any cross covariance between auto and cross power spectra.

We include errors corresponding to the quadrature sum of three contributions: the cosmic variance associated to biased tracers in a ``Fixed \& Paired'' simulation with our box size, the Poisson noise associated to the considered number density of biased tracers and an extra error corresponding to the 1\% of the power spectrum signal at each scale. For the former, we use the expressions derived in \cite{MaionAnguloZennaro2022}, weighting each contribution by a rough estimate of the corresponding bias parameter. For the Poisson contribution, for a sample with number density $\bar{n}$, we compute
\begin{equation}
    \sigma_{\rm Poisson}^2 = \dfrac{2}{N_k} \dfrac{1}{\bar{n}^2},
\end{equation}
where $N_k = V / (2\pi)^3 4 \pi k^2 \dif k$ is the number of wavemodes falling in each $k$ bin. Finally, we find the latter contribution to the error to be required to account for other sources of noise that are not captured in the ``Fixed \& Paired'' predictions.

We sample the posterior probability
\begin{equation}
    p[\boldsymbol{\vartheta}| \tilde{P}(k)] = \dfrac{p[\tilde{P}(k) | \boldsymbol{\vartheta}] p[\boldsymbol{\vartheta}]}{p[\tilde{P}(k)]},
\end{equation}
using the optimized simultaneous ellipsoidal nested sampler algorithm \textsc{MultiNest} \citep{FerozHobsonBridges2008} with its python wrapper \texttt{pymultinest} \citep{BuchnerEtal2014}. The sampling is considered converged when the precision on the log-evidence determined by \textsc{MultiNest} falls below 0.5 dex. Upon visual inspection of the results, we discard the initial 65\% of each chain as burn-in, and estimate our 2D and 1D projections of the posterior with the remaining points of each run.


\begin{figure}
    \includegraphics[width=0.48\textwidth]{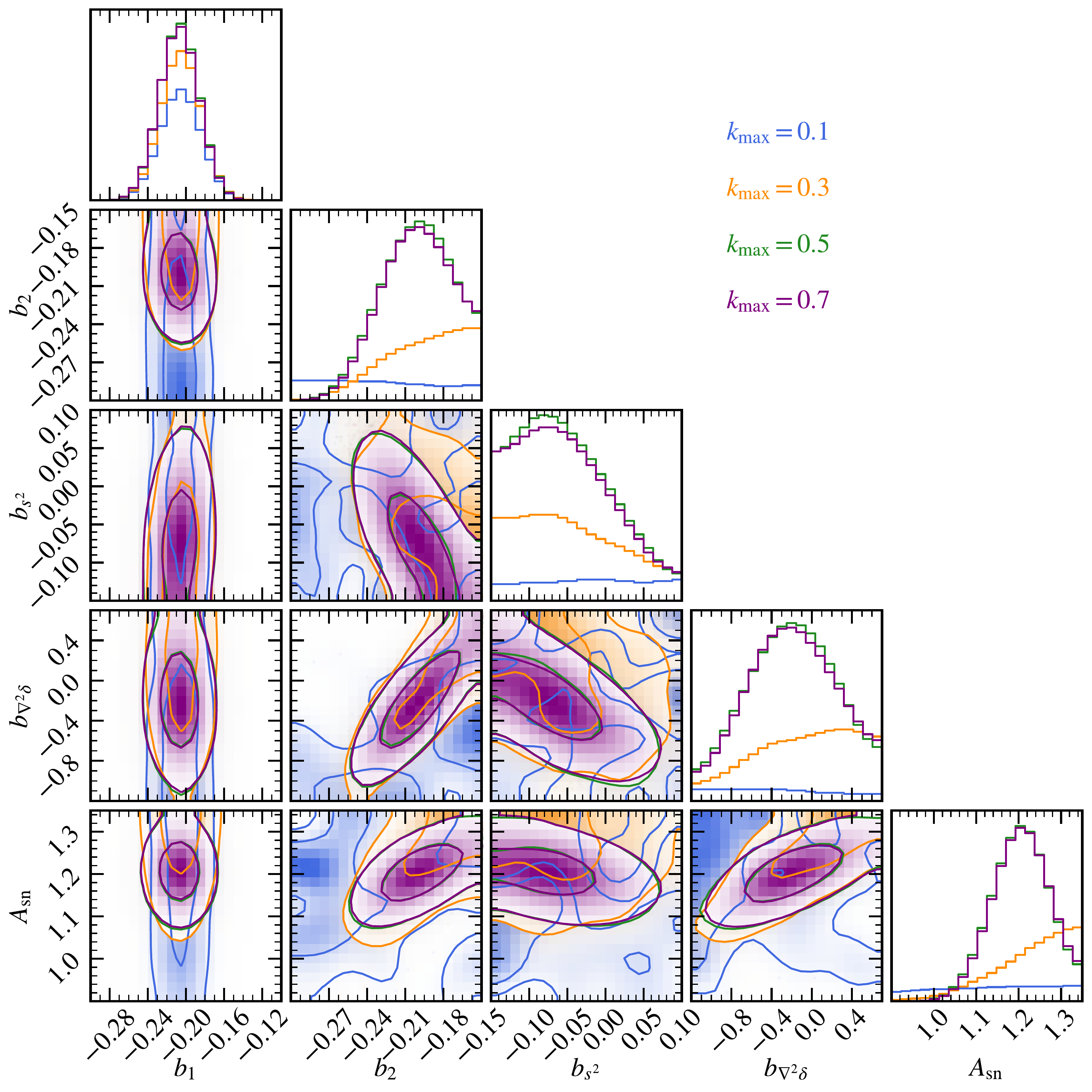}
    \caption{The 1-$\sigma$ and 2-$\sigma$ contours corresponding to the parameters fitted to our fiducial SFR-selected galaxy sample at $z=0$ with number density $n=10^{-3} h^3 \mathrm{Mpc}^{-3}$. The model corresponding to the best fitting parameters is shown in Fig. \ref{fig::bestfit-model}. Different colours mark different values of $k_{\rm max}$, from $0.1$ to $0.7 \ihMpc$. The Lagrangian fields used for the model are smoothed at $k_{\rm d} = 0.75 \ihMpc$.}
    \label{fig::bestfit-model-contours}
\end{figure}

In Fig. \ref{fig::bestfit-model-contours} we present an example of the fitting procedure applied to a reference case of SFR-selected galaxies at $z=0$. This sample, from the {\nenya} simulation, has a number density of $\bar{n} = 0.001 h^3 \mathrm{Mpc}^{-3}$ and assumes the SHAMe parameters found to reproduce the SFR-selected galaxy clustering of the TNG-300 simulation, taken from Tab. 1 of \cite{ContrerasAnguloZennaro2021b}. In particular, Fig. \ref{fig::bestfit-model-contours} shows the posterior obtained fitting the galaxy power spectrum including progressively smaller scales. Using $k_{\rm max} = {0.1, 0.3, 0.5, 0.7} \ihMpc$ we find that, while the contours shrink as $k_{\rm max}$ increases, the best fitting parameters remain compatible along all this range of scales.

\begin{figure}
    \includegraphics[width=0.48\textwidth]{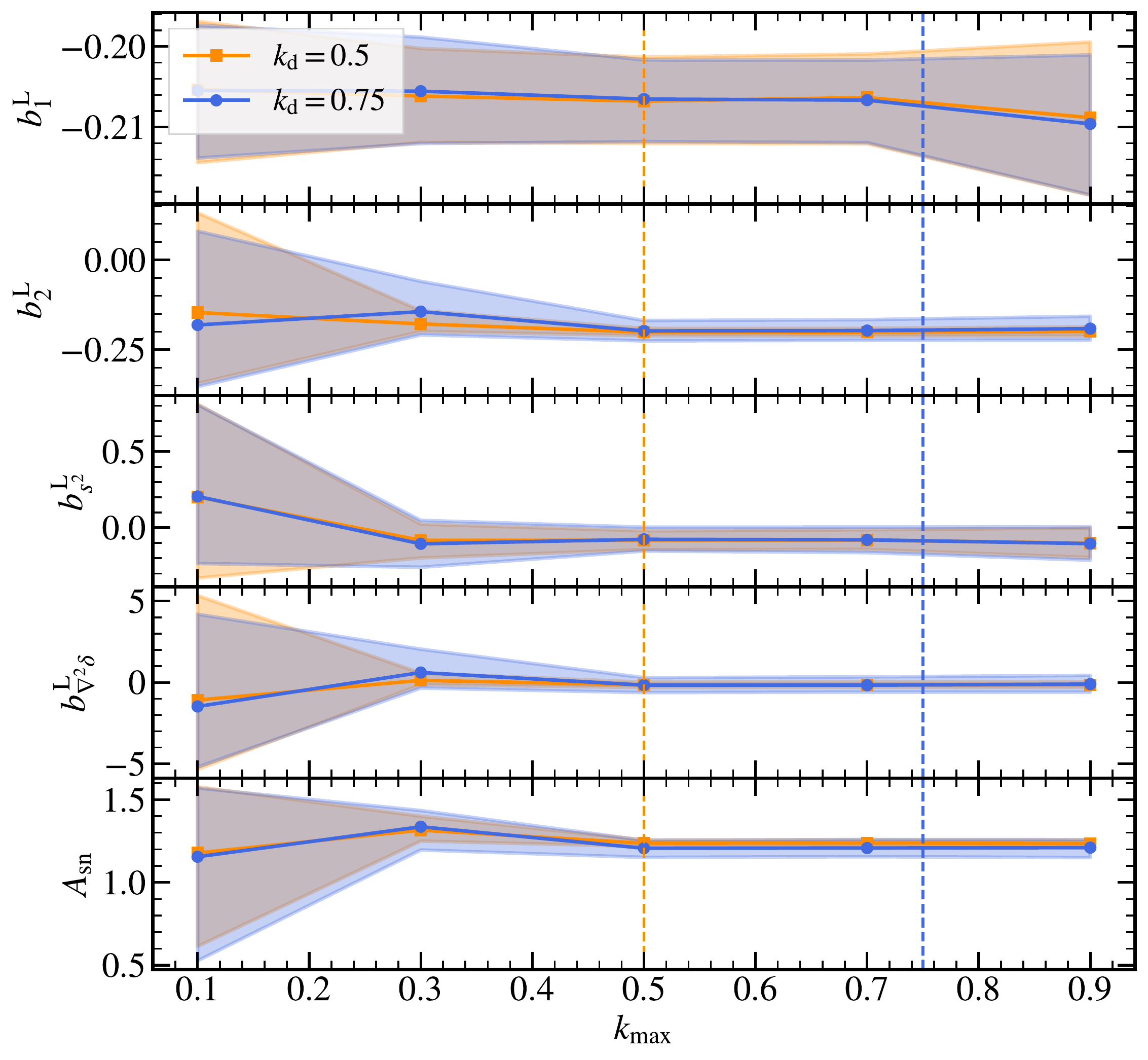}
    \caption{The values of the free parameters of the model fitted to the same data as in Fig. \ref{fig::bestfit-model-contours} as we vary the minimum scale included in the fit. Blue lines refer to smoothing the Lagrangian fields at a scale $k_{\rm d} = 0.75 \ihMpc$, while orange lines represent $k_{\rm d} = 0.5 \ihMpc$. The shaded areas represent the 1-$\sigma$ credibility level inferred from the nested sampling chains.}
    \label{fig::bestfit-kmax}
\end{figure}

To further investigate the optimal configuration to adopt for our fitting procedure, we present in Fig. \ref{fig::bestfit-kmax} the dependence of our best fitting bias parameters on $k_{\rm max}$, this time additionally focussing on the difference induced by choosing a different smoothing scale for the Lagrangian fields used to compute the model. In particular, besides our fiducial value of $k_{\rm d} = 0.75 \ihMpc$, we also consider the case of $k_{\rm d} = 0.5 \ihMpc$. We find that our best fitting parameters are robustly constrained as we vary $k_{\rm max}$, showing no significant scale dependence all the way down to the smaller scale allowed by our smoothing, namely $k_{\rm max} = 0.7 \ihMpc$. Our results are also not strongly affected by the smoothing scale we adopt in the model.

Therefore, we limit our fits to $k_{\rm max}$ values corresponding to scales larger than the smoothing scale, since we expect the model to eventually break down on smaller scales. We choose as our damping scale $k_{\rm d} = 0.75 \ihMpc$, to be consistent with the choice of \cite{ZennaroEtal2021}. Therefore, we will limit our fits to $k_{\rm max} = 0.7 \ihMpc$ in the remainder of this work.

However, we also enforce a limitation of the scales included in the fit. To ensure that the signal of the galaxy auto power spectrum is not dominated by shot-noise, we define a shot-noise dependent $k_{\rm sn, max}$, corresponding to the scale at which the power spectrum is 1.5 times the level of the poisson noise. We then consider $k_{\rm max} = \min(0.7, k_{\rm sn, max})$. There are two reasons for this: the first one is that there is no additional information to be extracted from shot-noise dominated scales; the second one is that the shot noise model considered is not accurate on the transition between signal-dominated and noise-dominated scales. We explicitly checked that this extra 1.5-factor is enough to ensure that we always consider signal-dominated scales. However different in the details of the implementation, this approach is similar to the one adopted in \cite{BarreiraLazeyrasSchmidt2021}, where the authors compute a phase-correlation coefficient between galaxies and matter to establish the scale at which the noise makes effectively meaningless the bias formalism.

\begin{figure}
    \includegraphics[width=0.48\textwidth]{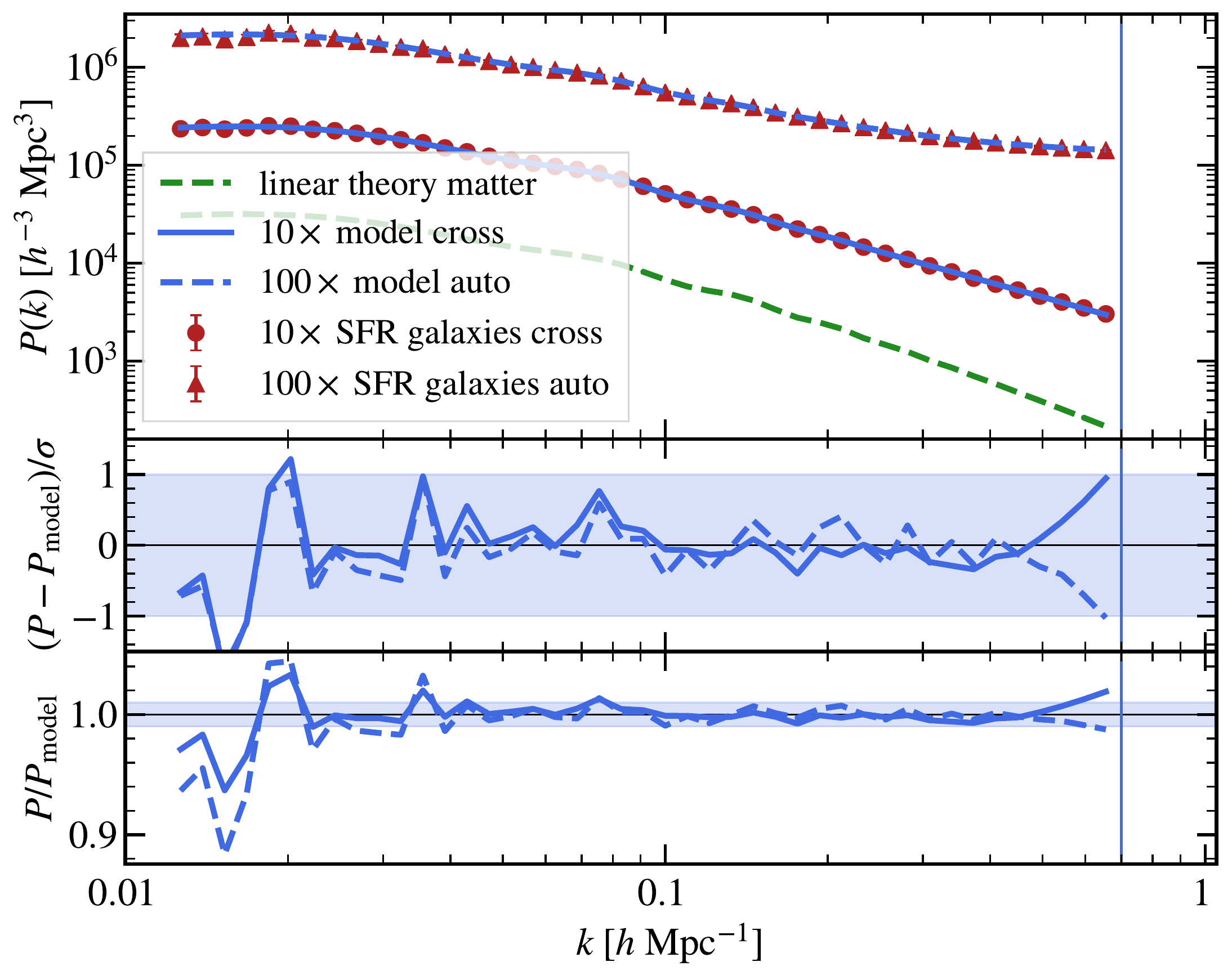}
    \caption{The galaxy auto and cross power spectrum for our fiducial SFR-selected sample with number density $n=10^{-3} h^3 \mathrm{Mpc}^{-3}$ at $z=0$ (red triangles and circles respectively), and our model computed with the best fitting set of parameters (blue solid lines). For reference, we also plot the linear theory prediction for the matter power spectrum as a green dashed line. The $\chi^2/\nu$ corresponding to this fit (with $81$ degrees of freedom) is $0.14$. The middle panel shows the difference between measures and best-fitting model in units of the error at each wavenumber, with dashed lines representing the galaxy auto power spectrum, and solid lines the galaxy-matter cross power spectrum. The same convention applies to the lower panel, showing the ratio between the measured spectra and the best-fitting model. The fit includes scales down to $k_{\rm max} = 0.7 \ihMpc$ (marked with a vertical line).}
    \label{fig::bestfit-model}
\end{figure}

Fig. \ref{fig::bestfit-model} shows the best fitting model for our SFR-selected galaxies down to $k_{\rm max}=0.7 \ihMpc$. The model describes the measured auto and cross galaxy power spectrum within the considered error bars, corresponding to a percent level agreement.

While we have shown this analysis with our fiducial sample of SFR-selected galaxies, we have explicitly checked that these value of $k_{\rm d}$ and $k_{\rm max}$ are well suited also for analysing SM-selected galaxies. Moreover, we have repeated these tests for a population of haloes as well.

\begin{figure}
    \includegraphics[width=0.48\textwidth]{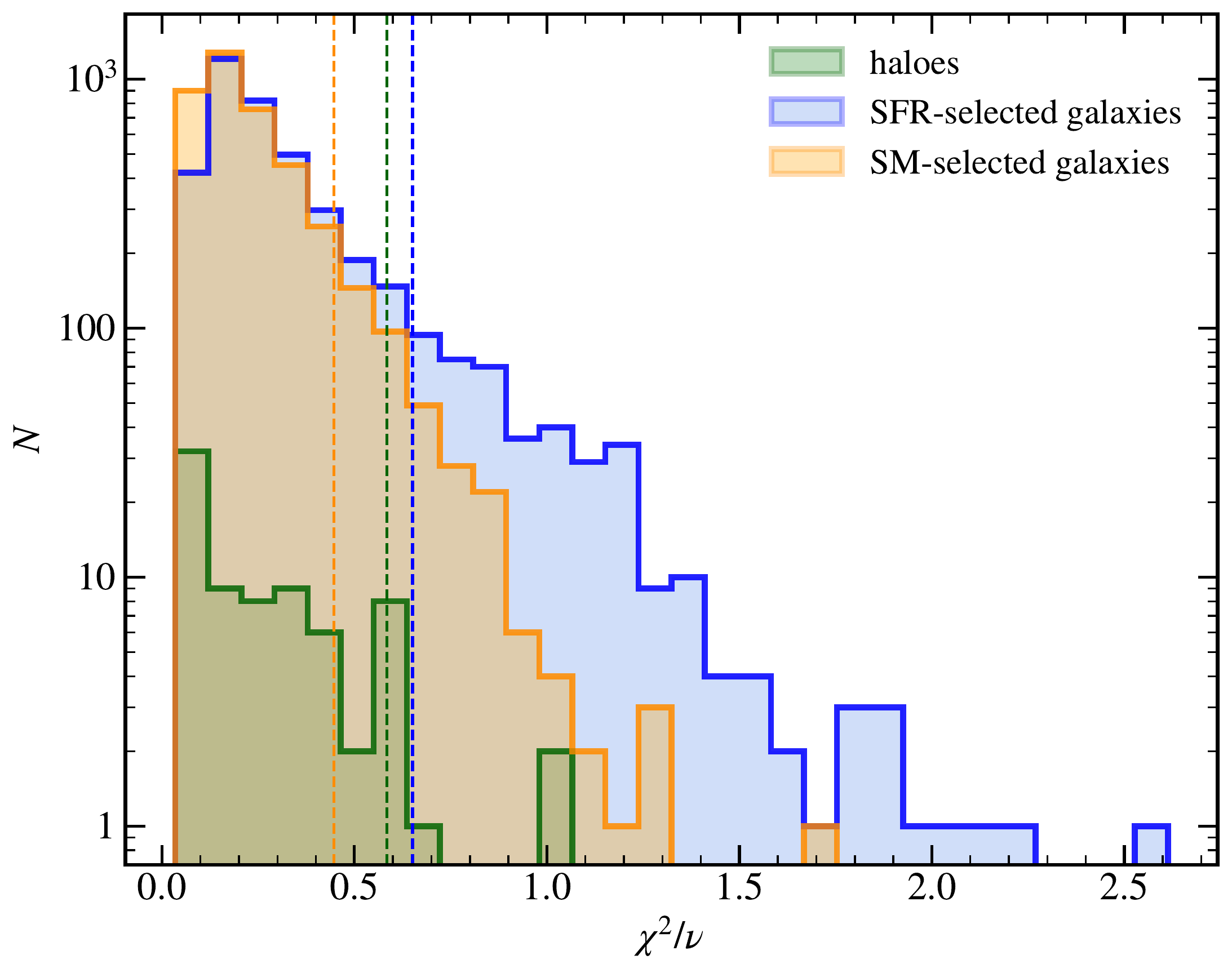}
    \caption{The distribution of reduced $\chi^2$ values for the best-fitting parameters. The vertical lines mark the $0.9$ quantile of each distribution. In general, we fit 50 $k$-bins for the auto power spectrum and $50$ for the cross spectrum, with $5$ free parameters, corresponding to $95$ degrees of freedom; however, after enforcing our $k_{\rm sn,max}$ criterion, the exact number of degrees of freedom can vary for some of the samples.}
    \label{fig::chi2s}
\end{figure}

Finally, to check that all our fits resulted in sensible bias parameters, we show in Fig. \ref{fig::chi2s} the distribution of the values of the reduced $\chi^2$ of each fitted bias set. The vertical dashed lines mark the $q=0.9$ percentile of each distribution. We remind the reader that our data vector is composed of a measurement of the galaxy auto power spectrum (in 50 $k$-bins), one measurement of the galaxy-matter cross power spectrum (also in $50$ $k$-bins), while we leave 5 parameters free. This would results in $95$ degrees of freedom. However, after imposing cuts in $k_{\rm max}$ and $k_{\rm sn, max}$, we are left with fewer degrees of freedom, the exact amount being different for each model.

Considering the $q=0.9$ quantile, 90\% of our reduced $\chi^2$ values fall below $\sim 0.6$ for haloes and SM-selected galaxies, and below $\sim 0.7$ for SFR-selected galaxies. Please note that these values of reduced $\chi^2$ below $1$ do not necessarily indicate any problems with our computation of the errors, but are justified by the fact that our data and model are not independent, but are both drawn from the same simulations.

As a consequence, we conclude that the hybrid Lagrangian bias expansion model in real space is very well suited to describe the auto and cross power spectra of biased tracers, including extremely small scales ($k_{\rm max} = 0.7 \ihMpc$) as long as they are not dominated by shot noise. We remark that halo and galaxy samples with extremely different bias properties have been considered.

\section{Results}\label{sec::results}

In this section, we will present the bias parameters obtained from fitting a variety of mass tracers. In particular, we will first test the reliability of our model and fitting procedure with haloes, comparing our results to the findings of a number of previous works. We will then proceed to separately present the same bias relations obtained for SM-selected and SFR-selected galaxies at $z=0$. Finally, we will summarise our results comparing, at the same time, all tracers, cosmologies, number densities, and redshifts.

\subsection{The bias-mass relation for haloes}\label{subsec:halo-bm}
Before focusing on galaxies, we begin by using our model to infer the bias parameters of populations of haloes of different masses. In this case, a number of approaches are already commonly used to make predictions. Depending on the bias parameter considered, predictions can be obtained both from theories and from fitting functions typically calibrated with haloes in high-resolution simulations.

Lagrangian bias parameters of arbitrary order can be predicted in the context of the peak-background split (PBS) formalism \citep[see review in][and references therein]{DesjacquesJeongSchmidt2016}. This means that, for a given mass function and background density, the $n$-th order derivative of that mass function can be related to the $n$-th order Lagrangian bias parameter through
\begin{equation}
    b_N^{\rm L}(M) = \left[\dfrac{-1}{\sigma(R(M))}\right]^N \dfrac{1}{\nu f(\nu)} \dfrac{\dif^N [\nu f(\nu)]}{\dif \nu^N},
    \label{eq::pbs}
\end{equation}
where $f(\nu)$ is an analytic parametrisation of the halo mass function and $\nu \equiv \delta_{\rm c}(a) / \sigma(M,a)$ is the peak height. Here, $\delta_{\rm c}(a)$ is the time-dependent threshold for the collapse of matter overdensities. We use the value obtained assuming spherical collapse, which, albeit somewhat simplistic, is accurate enough for our case. The denominator comes from $\sigma^2(R)$, the variance of the linear density field smoothed on a scale $R$, such that spheres of radius $R$ would on average contain the mass $M$. For our predictions, we employ the halo mass function fit proposed by \cite{Ondaro-MalleaEtal2021}. Please note that, unlike other proposed halo mass function models, the one of \cite{Ondaro-MalleaEtal2021} explicitly accounts for the non-universality of the mass functions and, therefore, it is in principle cosmology and redshift dependent. We disregard this important feature here, plotting only the predictions obtained for one cosmology at one redshift (\nenya\ at $z=0$).

For the linear bias parameter $b_1^{\rm L}(M)$ we obtain predictions using the PBS approach (Eq. \ref{eq::pbs}), and using the fitting function presented in \cite{TinkerEtal2010}. This fitting function predicts the value of the Eulerian linear bias, which we convert to Lagrangian adopting $b_1^{\rm L} = b_1 - 1$.

For the quadratic bias parameter $b_2^{\rm L}(M)$ we employ predictions obtained with the PBS (Eq. \ref{eq::pbs}). Moreover, we compare to the fitting formula presented in \cite{LazeyrasEtal2016}. Note that in this paper a fit to the relation between the Eulerian bias parameters $b_2(b_1)$ is presented, with $b_2(b_1) = 0.412 - 2.143 b_1 + 0.929 b_1^2 + 0.008 b_1^3$. As a consequence, first we use the fitting function of \cite{TinkerEtal2010} to obtain $b_1(M)$, which we use to obtain $b_2(b_1)$. Finally, we convert the Eulerian squared bias parameter to its Lagrangian counterpart adopting the (approximated) relation $b_2^{L} = [b_2 - 8/21 \, (b_1 - 1)] / 2$, as in \cite{ShethChanScoccimarro2013} (note the factor two difference from the published formula to account for the different definition of $b_2$).

The tidal bias parameter can be predicted, in the context of the ``Lagrangian Local In Matter Density'' (LLIMD) by assuming it is exactly zero, $b_{s^2}^{\rm L} = 0$. This is an approximation expected to break down. In particular, in \cite{ModiEtal2017} the authors showed that, indeed, they found a nonzero signal for the Lagrangian tidal bias, and proposed a fitting function for $b_{s^2}(M)$. A subsequent work by \cite{LazeyrasSchmidt2018} could not confirm the results of \cite{ModiEtal2017}, but still found a slight departure from zero. We consider the fitting formula of \cite{ModiEtal2017} for $b_{s^2}(b_1)$. Moreover, we consider the prediction of $b_{s^2}(b_1)$ obtained from applying the excursion set formalism \citep{ShethChanScoccimarro2013}, $b_{s^2}(b_1) = 0.524 - 0.547 b_1 + 0.046 b_1^2$. Both formulas predict the tidal bias in Eulerian space, which we convert to its Lagrangian counterpart assuming $b_{s^2}^{\rm L} = b_{s^2} + 2/7 \, (b_{1} - 1)$, \citep{DesjacquesJeongSchmidt2016}.

Finally, for $b_{\nabla^2\delta}^{\rm L}$ we can think that it scales as the squared Lagrangian radius of the tracers, $b_{\nabla^2\delta^2}^{\rm L} \sim -2 R_{\rm L}^2$ \citep[see, for example][and references therein for a review of different models for the Eulerian higher-derivative bias parameter]{LazeyrasSchmidt2019}. Another prediction (known to break down for lower mass haloes) comes from the so-called peak theory; in this context \citep{EliaEtal2012,BaldaufDesjacquesSeljak2015}, the scale dependent peak bias $b_{01}(\nu)$ is expected to be the driving contribution to $b_{\nabla^2\delta}^{\rm L}$. Therefore we use the approximation $b_{\nabla^2\delta}^{\rm L} \sim -b_{01} $. This prediction is already referred to Lagrangian bias parameters, the Eulerian counterpart being impossible to obtain without assuming a model for velocity bias.

For each cosmology and redshift of our simulation set, we select haloes by splitting them into 10 logarithmically spaced mass bins, spanning $10^{10} < M_{\rm 200c} / [h^{-1} \mathrm{M}_\odot] < 10^{15}$. Since we are dealing with very different cosmologies and redshifts and the bias function is only supposed to be universal when expressed in terms of peak height, we convert our masses into the peak height $\nu$.

\begin{figure*}
    \includegraphics[width=\textwidth]{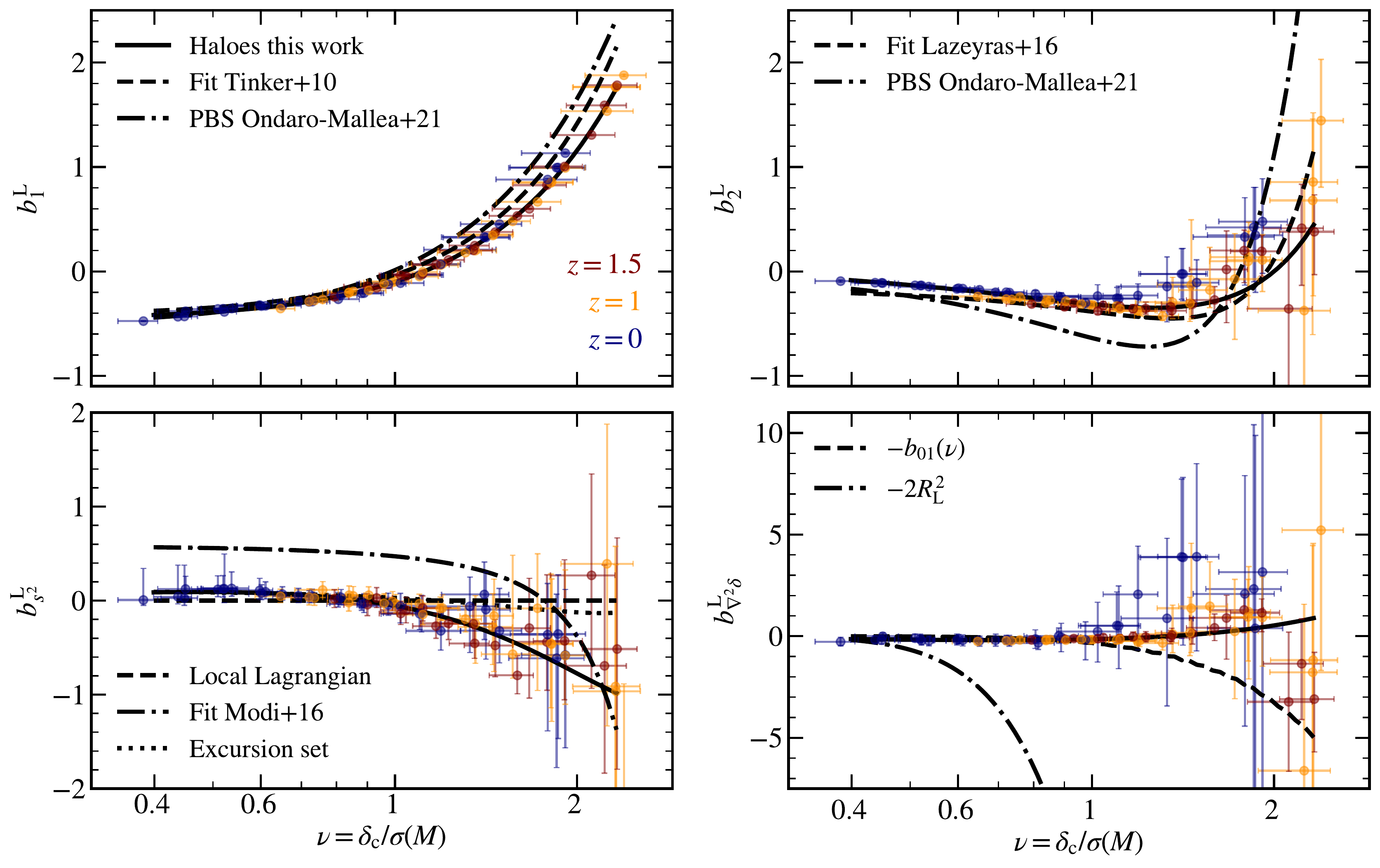}
    \caption{The Lagrangian bias parameters that best fit the halo auto and cross power spectrum in 10 mass bins at redshifts $z=0, 1$ and $1.5$, for our four cosmologies. Fits include scales $k < \min(0.7 \ihMpc, k_{\rm sn,max})$ and always assume a smoothing scale $k_{\rm d} = 0.75\ihMpc$. Points colored in blue correspond to $z=0$, orange points correspond to $z=1$ and red ones to $z=1.5$. Black solid lines show our best fitting functions, while non-solid lines show predictions available in the literature for the bias-mass relations.}
    \label{fig::halo-mass}
\end{figure*}

Fig. \ref{fig::halo-mass} shows the marginalized bias parameters that best fit our halo data. To each point we associate an error bar along the x-axis, marking the width of the mass bin once transformed into peak height. Moreover, we plot the $68\%$ CL of the marginalized pdf as an error bar along the y axis.

For all bias parameters, the $b_i$-$\nu$ relation is confirmed to be universal to very good degree, considering how tightly they describe a univocal relation although coming from 4 different cosmologies at 2 different redshifts. Moreover, the value of the bias parameters we obtain are in very good agreement with all the predictions considered. Some discrepancies appear in the best fitting values of $b_2$ at $\nu > 1$. However, these are still compatible with all the considered predictions within $1$-$\sigma$. We note that these discrepancies can be originated by different factors. One is that the actual value of all bias parameters (and of quadratic bias parameters in particular) depends on the smoothing performed on the original density field, especially for extreme choices of the smoothing scales (way different from the difference considered in this work in Fig. \ref{fig::bestfit-kmax}). A second reason is that we do not expect that bias parameters in our model (including fairly small scales and not including resummation terms) exactly coincide with the large-scale limit typically quoted in the literature. Finally, because comparisons between Lagrangian and Eluerian bias parameters rely on approximated formulas connecting the two different formalisms. Nonetheless, we stress that such differences (whose exact origin would be interesting to investigate) are not statistically significative. We use this result to support the reliability of the bias parameters we infer using our model.

\subsection{Bias relations}
We now present the relation of the higher order bias parameters with the linear bias. We do so focusing first on haloes, and then on SM-selected and SFR-selected galaxies. For galaxies, we start by analysing our results at $z=0$, exploring possible correlations of bias parameters with the galaxy formation parameters. Finally, we then combine all of our halo and galaxy catalogues (also including those at $z=1$).

\subsubsection{Coevolution relations for haloes}
\begin{figure}
    \includegraphics[width=0.48\textwidth]{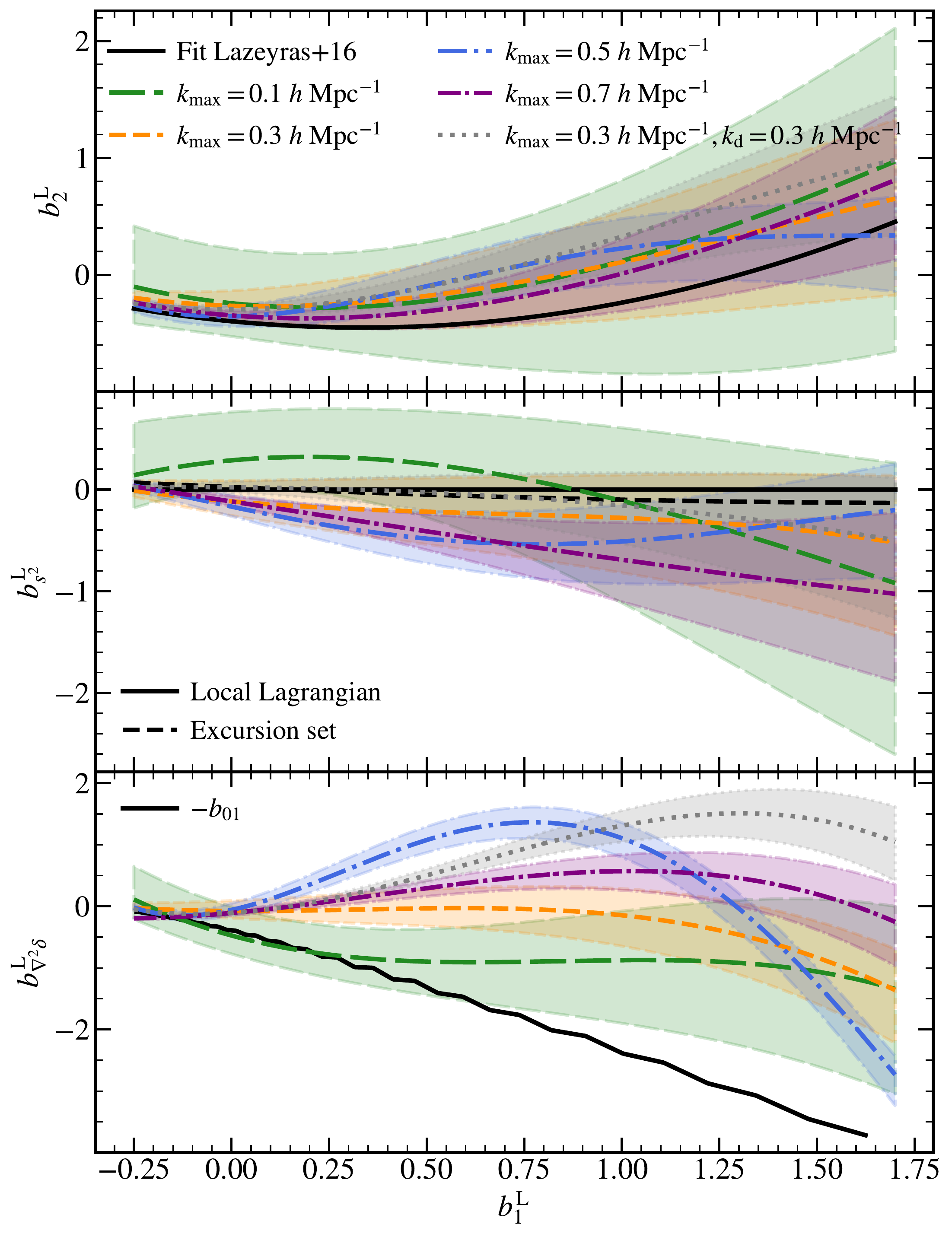}
    \caption{The Lagrangian bias relations for haloes selected in 10 mass bins in $10^{10} < M_{\rm 200c} / [h^{-1} \mathrm{M}_\odot] < 10^{15}$. Different colours refer to the same bias relations obtained fitting the halo auto and cross power spectra up to $k_{\rm max} = [0.1, 0.3, 0.5, 0.7] \ihMpc$. In all cases, we keep our smoothing fixed to $k_{\rm d} = 0.75 \ihMpc$. Shaded areas have been obtained fitting a polynomial to our best fitting parameters increased or decreased by their corresponding marginalized 1-$\sigma$ C.L.. Black lines represent fitting formulae and predictions available in the literature, as described in Section \ref{subsec:halo-bm}.}
    \label{fig::kmax-halos}
\end{figure}

We present in Fig. \ref{fig::kmax-halos} the relations between Lagrangian bias parameters that we obtain considering our halo samples. In particular, for clarity, we show here polynomial fitting functions that capture the mean and 1-$\sigma$ dispersion of our best fitting bias parameters. We repeat here some state-of-the-art relations available in the literature for comparison, including the $b_2(b_1)$ fit from \cite{LazeyrasEtal2016}, the LLIMD and excursion set predictions for $b_{s^2}^{\rm L}$ and the scale dependent peak bias $b_{01}$ for $b_{\nabla^2\delta}^{\rm L}$. All these relations have been presented in Section \ref{subsec:halo-bm}.

The different lines in Fig. \ref{fig::kmax-halos} correspond to different choices of $k_{\rm max}$, spanning the interval $[0.1, 0.7] \, \ihMpc$. All these cases share the same smoothing scale $k_{\rm d} = 0.75\,\ihMpc$. Moreover, we also show one case with different smoothing, namely $k_{\rm d} = 0.3\,\ihMpc$; in this case we assume $k_{\rm max} = 0.3 \, \ihMpc$. We can see that for both the $b_2^{\rm L}(b_1^{\rm L})$ and $b_{s^2}^{\rm L}(b_1^{\rm L})$ relations we do not find any significant dependence on $k_{\rm max}$ nor $k_{\rm d}$, the main difference being larger scatter for lower $k_{\rm max}$ values. This is compatible with the fact that for low values of $k_{\rm max}$ the free parameters of our model become more and more unconstrained (see also Figs. \ref{fig::bestfit-model-contours} and \ref{fig::bestfit-kmax}). On the contrary, the $b_{\nabla^2\delta}^{\rm L}(b_1^{\rm L})$ relation seems to depend more heavily on the choice of $k_{\rm max}$ and smoothing scale.

Finally, consistently with our findings in Fig. \ref{fig::halo-mass}, we find that our fits systematically describe a slightly different $b_2^{\rm L}(b_1^{\rm L})$ relation when compared to the fitting function of \cite{LazeyrasEtal2016}, even though the two are compatible within the given errors. Once again, we do not expect our bias values to coincide with the corresponding large-scale bias parameters presented in that work. We also find that our haloes present $b_{s^2}^{\rm L}(b_1^{\rm L})$ relations roughly compatible with both the LLIMD approximation and the excursion set prediction. Lastly, we notice that for the $b_{\nabla^2\delta}^{\rm L}(b_1^{\rm L})$ relation, only considering large scales produces results closer to the prediction of $b_{01}$. We conclude that it is important to consistently choose smoothing scale and fit limits in order to be able to compare results. In the remainder of this paper, we will always compare to the coevolution relations of haloes with $k_{\rm max} = 0.7 \, \ihMpc$ and $k_{\rm d} = 0.75 \, \ihMpc$. In Table \ref{tab:polyfits} we present the polynomial fitting function corresponding to this case.

\subsubsection{SM-selected galaxies at $z=0$}
\begin{figure*}
    \includegraphics[width=\textwidth]{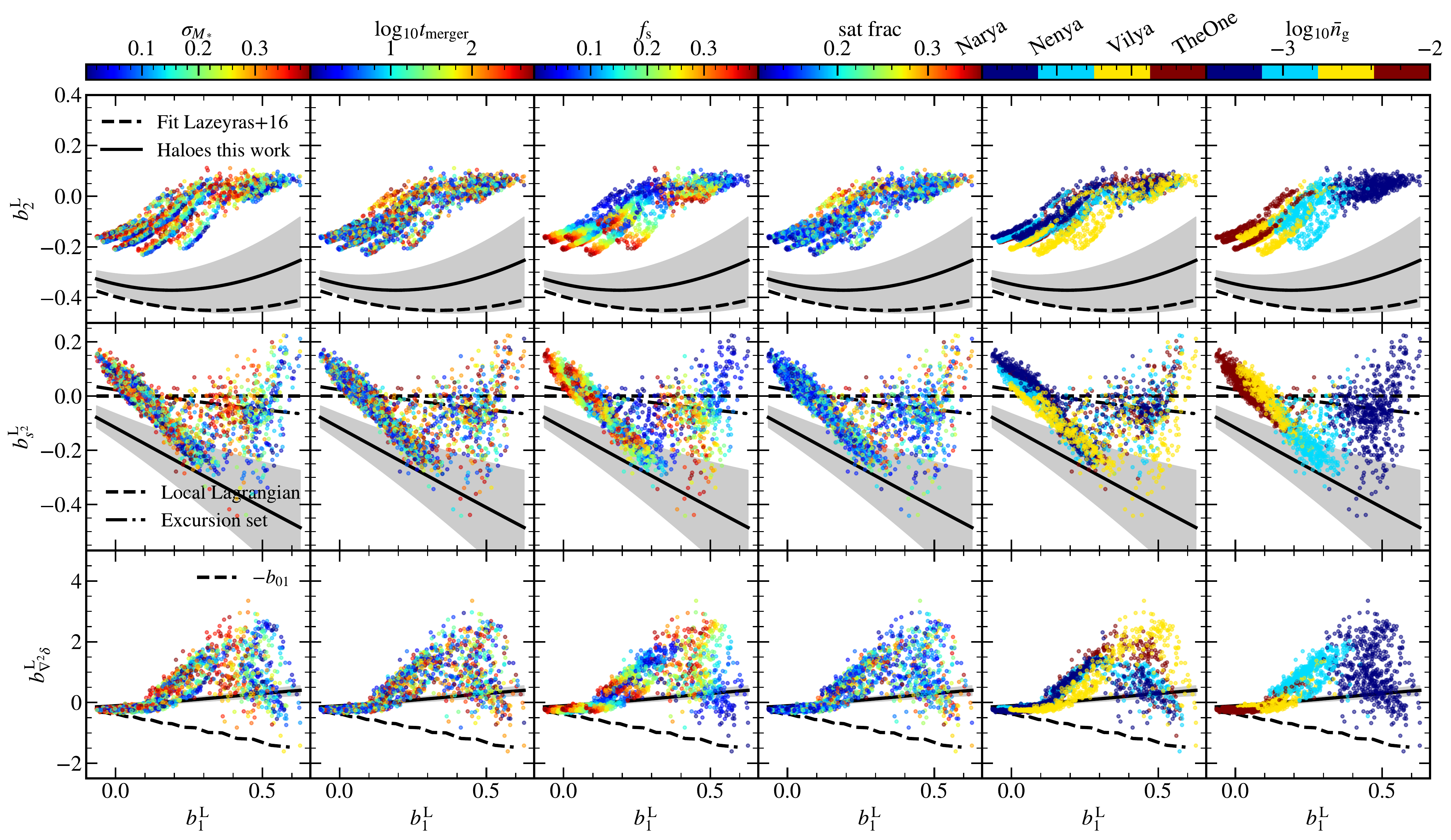}
    \caption{The relations between higher order lagrangian bias parameters and $b_1^{\rm L}$ for SM selected galaxies at $z=0$, with number densities from $3 \times 10^{-4}$ to $0.01 \,\, h^3 \, \mathrm{Mpc}^{-3}$. Each row corresponds to one bias parameter ($b_2^{\rm L}$, $b_{s^2}^{\rm L}$, $b_{\nabla^2\delta}^{\rm L}$), while each column shows the $b_i^{\rm L} - b_1^{\rm L}$ relation color coded according to one of the free parameters of the galaxy formation model. Specifically, on the left colors from blue to red correspond to increasing values of $\sigma_{M_*}$, in the second column they correspond to increasing values of $t_{\rm merger}$, and in the middle column to increasing values of the tidal stripping parameter $f_{\rm s}$. The three rightmost columns show points colour-coded according to satellite fraction, cosmology, and number density. Fits reported in this plot consider $k_{\rm max} = \min(0.7 \ihMpc, k_{\rm sn,max})$. Black lines show predictions obtained for these quantities from the literature and from haloes of our simulations.}
    \label{fig::b1-b2-sm}
\end{figure*}

We now move to investigate these coevolution relations for samples of galaxies. In Fig. \ref{fig::b1-b2-sm} we show the bias relations obtained for our sample of SM selected galaxies at redshift $z=0$, for all four cosmologies considered. In particular, we focus on the relation of each higher order bias parameter with $b_1$. We repeat our results several times, color coding each point according to a different property. In the first three columns, we color code each point based on the three free parameters of the galaxy formation model for SM-selected galaxies, thus addressing the possible dependence of the Lagrangian bias relations on the galaxy model. In the following columns, we assign colors according to the satellite fraction of the sample, cosmology, and, finally, the number density.

First of all, we notice that, at a fixed cosmology, each of the higher order bias parameters exhibits a quite tight correlation with $b_1$ up to $b_1 \lesssim 0.2 - 0.4$. For larger values of $b_1$, the relations are more scattered. This more scattered behaviour corresponds to samples with lowest number density, where higher order bias parameters are determined with the least precision. Moreover, we do not find a strong dependence of the bias relations on $\sigma_{M_*}$ and $t_{\rm merger}$, even if $\sigma_{M_*}$ seems to slightly anti-correlate with $b_1$ at fixed cosmology. We notice that higher values of $f_s$ correlate with low $b_1$, low $b_2$, and $b_{s^2} > 0$. Galaxy samples with high $f_s$ destroy relatively early their satellite galaxies (as also shown by the fourth column in Fig. \ref{fig::b1-b2-sm}), which is compatible with a lower linear bias. Finally, we find a small, residual correlation of these bias relations with cosmology.

In addition, in Fig. \ref{fig::b1-b2-sm} we also show a selection of predictions for the $b_i(b_1)$ relations. In particular, we display the fit to the $b_2(b_1)$ relation presented in \cite{LazeyrasEtal2016} and adapted to Lagrangian bias parameters as described in Section \ref{subsec:halo-bm}; a fit to the haloes of section \ref{subsec:halo-bm} using the polynomial presented in Table \ref{tab:polyfits}, surrounded by a 1-$\sigma$ region based on our dataset; the local Lagrangian prediction for the tidal bias parameter; the excursion set prediction for $b_{s^2}^{\rm L}(b_1^{\rm L})$ presented in Section \ref{subsec:halo-bm}; and $b_{\nabla^2\delta} = - b_{01}$. We notice that the bias parameters from this galaxy sample are shifted away from the considered predictions. In particular, $b_2^{\rm L}$, at fixed $b_1^{\rm L}$, is consistently higher than both the Lazeyras fitting function, and the fitting function developed in this work. The tidal bias parameters shows a large scatter around both the local Lagrangian prediction and the prediction from excursion set theory. Finally, the laplacian bias parameter shows a departure from the prediction from peak theory at $b_1^{\rm L} > 0.2$. One important caveat here is that (for the sake of clarity) we are not plotting error bars in Fig. \ref{fig::b1-b2-sm}; we will show error bars in the summary plot at the end of this section.

\subsubsection{SFR-selected galaxies at $z=0$}
\begin{figure*}
    \includegraphics[width=\textwidth]{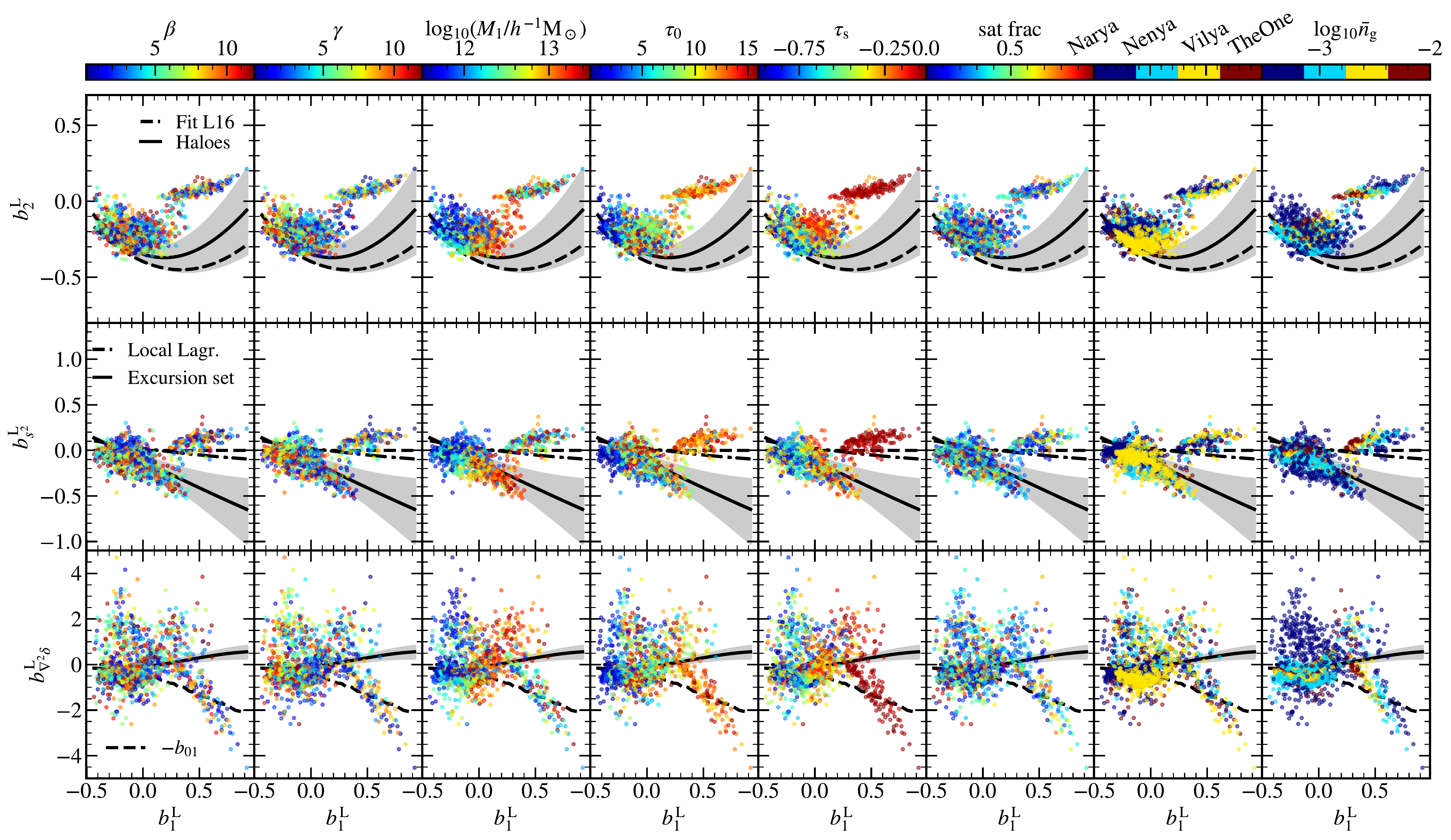}
    \caption{The relations between higher order lagrangian bias parameters and $b_1^{\rm L}$ for SFR selected galaxies at $z=0$, with number densities from $3 \times 10^{-4}$ to $0.01 \,\, h^3 \, \mathrm{Mpc}^{-3}$. Each row corresponds to one bias parameter ($b_2^{\rm L}$, $b_{s^2}^{\rm L}$, $b_{\nabla^2\delta}^{\rm L}$), while each column shows the $b_i^{\rm L} - b_1^{\rm L}$ relation color coded according to one of the free parameters of the galaxy formation model. Specifically, in the first five columns from the left, colors from blue to red correspond to increasing values of $\beta, \gamma, \log_{10}(M_1/h^{-1}\mathrm{M}_\odot), \tau_0,$ and $\tau_{\rm s}$. The three rightmost columns show points color-coded according to satellite fraction, cosmology, and number density. Fits reported in this plot consider $k_{\rm max} = \min(0.7 \ihMpc, k_{\rm sn,max})$. Black lines show predictions obtained for these quantities from the literature and from haloes of our simulations.}
    \label{fig::b1-b2-sfr}
\end{figure*}

Fig. \ref{fig::b1-b2-sfr} shows the relations between the bias parameters obtained from the SFR-selected galaxy samples, once again at $z=0$, and for all the number densities considered. Once again, the color of each point reflects the parameters of the galaxy formation model (first 5 columns), the satellite fraction of the sample (sixth column), and the cosmology and number density of the galaxy catalogue (rightmost columns).

In the case of SFR-selected galaxies, we find that the strongest correlations between galaxy formation parameters is with $\tau_0$ and $\tau_{\rm s}$, the parameter controlling the quenching of SFR. Low values of $\tau_{\rm 0}$ and $\tau_{\rm s}$ imply fast quenching, especially for galaxies living in low-mass hosts ($M_{h} < 10^{12} h^{-1} \, \mathrm{M}_{\sun}$). We expect these samples to be dominated by central galaxies, which is consistent with our finding that lower values of $\tau_{\rm 0}$ and $\tau_{\rm s}$ correlate with lower $b_{1}^{\rm L}$ and populate a region closer to the $b_2^{\rm L}(b_1^{\rm L})$ relation of haloes. Vice versa, large values of $\tau_{\rm 0}$ imply long quenching times, and large values of $\tau_{\rm s} \rightarrow 0$ imply that the quenching efficiency is independent of host mass. These samples, with higher satellite fraction, follow a $b_2^{\rm L}(b_1^{\rm L})$ relation more similar to that of SM selected galaxies.

The latter point is reflected also in the $b_{s^2}^{\rm L}(b_1^{\rm L})$ relation, which closely follows our fitting function calibrated with haloes for samples with fast quenching, and resembles more our results for SM selected galaxies for samples richer in satellites. Finally, we find that the $b_{\nabla^2\delta}^{\rm L}(b_1^{\rm L})$ relation for SFR-selected galaxies exhibits significantly more scatter than our fitting function obtained with haloes.

\subsubsection{All halo and galaxy samples}
\begin{figure}
    \includegraphics[width=0.48\textwidth]{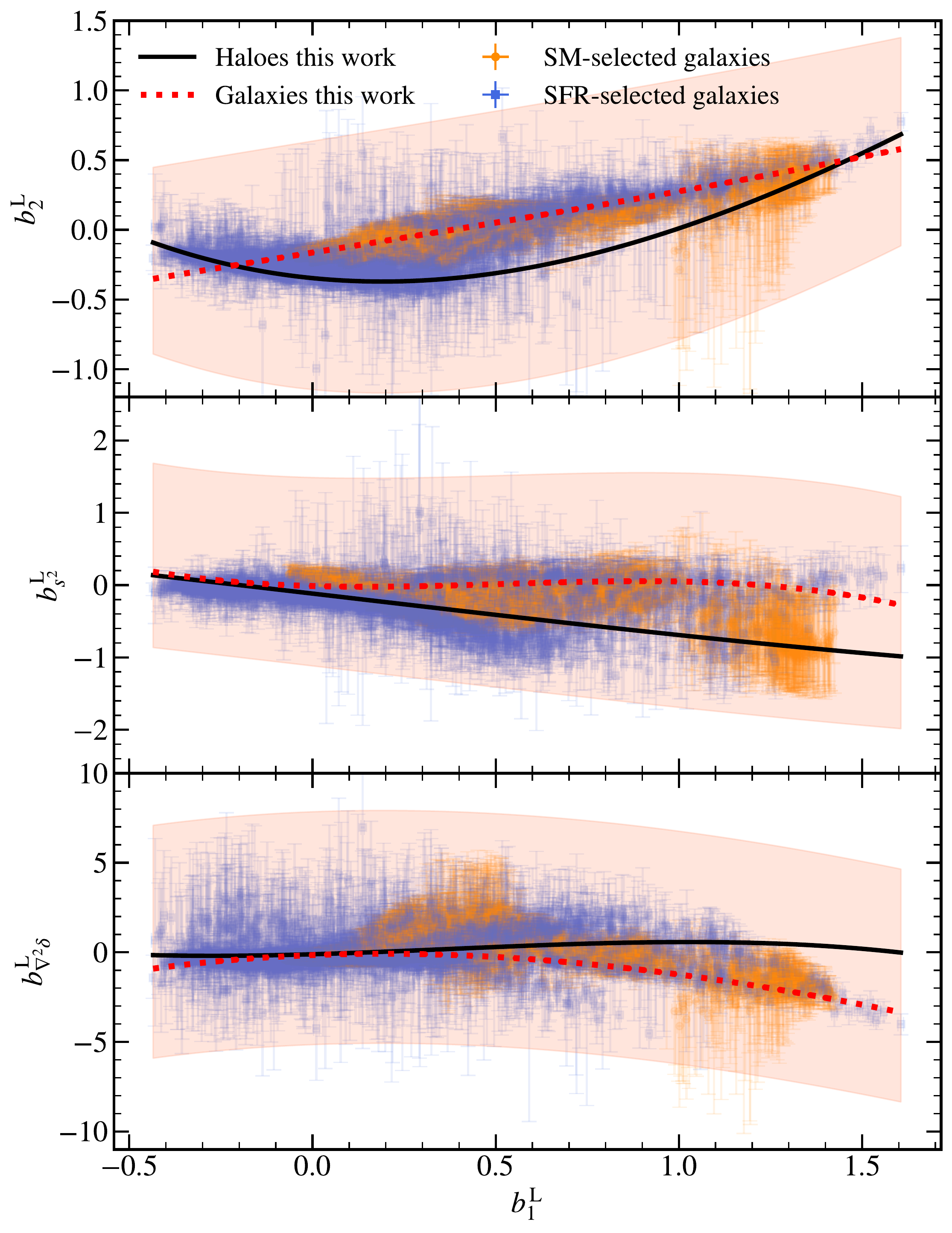}
    \caption{The relations between higher order lagrangian bias parameters and $b_1^{\rm L}$. Each row corresponds to one bias parameter ($b_2^{\rm L}$, $b_{s^2}^{\rm L}$, $b_{\nabla^2\delta}^{\rm L}$). Black lines show predictions obtained for these quantities from haloes of our simulations, while red lines refer to galaxies (see Tab. \ref{tab:polyfits}). We enclose all of our galaxy samples in a region given by Eq. (\ref{eq::priors}), here represented as a shaded red area.}
    \label{fig::b1-b2-summary}
\end{figure}

Fig. \ref{fig::b1-b2-summary} presents a summary view of the bias relations for the different biased tracers considered, including all number densities and both $z=0$ and $z=1$. In this figure we also show the error bars (corresponding to the $68\%$ C.L. from the marginalized posterior of each fit). Moreover, we include our fitting functions calibrated on haloes and a fitting function calibrated using all of our galaxy samples. Both formulae are presented in Table \ref{tab:polyfits}.

Especially in the $b_2^{\rm L}(b_1^{\rm L})$ case, we find a systematic shift of the galaxy relation from its halo counterpart. This happens not only when contrasting our galaxies with the fitting function from \cite{LazeyrasEtal2016} (which could exhibit differences due to different assumptions about the Lagrangian-Eulerian connection, smoothing scale, inclusion of small scales), but also when comparing with the fitting function calibrated with haloes from the catalogues developed for this work.

Finally, we enclose all of our galaxy samples in a hypervolume that will serve as prior knowledge for future Bayesian analyses. We define this hypervolume in terms of the halo and galaxy coevolution relations presented in Table \ref{tab:polyfits}, which here we call $b_{i,\mathrm{halo}}^{\rm L}(b_1^{\rm L})$ and $b_{i,\mathrm{gal}}^{\rm L}(b_1^{\rm L})$ respectively. For a given value of $b_1^{\rm L}$ the vertices of this hypervolume are given by
\begin{equation}
    \begin{split}
        b_2^{\rm L} = [b_{2, \mathrm{halo}}^{\rm L}(b_1^{\rm L}) - 0.8, \qquad b_{2, \mathrm{gal}}^{\rm L}(b_1^{\rm L}) + 0.8], \\
        b_{s^2}^{\rm L} = [b_{s^2, \mathrm{halo}}^{\rm L}(b_1^{\rm L}) - 1, \qquad b_{s^2, \mathrm{gal}}^{\rm L}(b_1^{\rm L}) + 1.5], \\
        b_{\nabla^2\delta}^{\rm L} = [b_{\nabla^2\delta, \mathrm{gal}}^{\rm L}(b_1^{\rm L}) - 5, \qquad b_{\nabla^2\delta, \mathrm{gal}}^{\rm L}(b_1^{\rm L}) + 8]. \\
    \end{split}
    \label{eq::priors}
\end{equation}
This region encloses 100\% of the galaxies in our sample and is represented with a shaded area in Fig. \ref{fig::b1-b2-summary}. An interesting point is that, even though we kept these relations quite loose (including 100\% of our galaxy samples, which feature very different and extreme galaxy formation model parameters and number densities), the resulting allowed regions are tighter than typical observational constraints on higher order bias parameters \citep[for example, in][using eBOSS emission line galaxies quadratic bias parameters are almost unconstrained and reflect the priors assumed in that analysis]{Ivanov2021}. Moreover we notice that, considering only galaxy samples whose parameters lie in a tight region ($1/8$ of the ranges reported in Table \ref{tab:LH}) centered around the best fitting parameters of the TNG300 simulation found in \cite{ContrerasAnguloZennaro2021b}, the coevolution relations obtained for galaxies do not become significantly tighter. Therefore we conclude that our choice of latin hypercube boundaries does not affect the allowed regions individuated in Eq. (\ref{eq::priors}).

We would like to spend a word of caution on the $b_{\nabla^2\delta}^{\rm L}(b_1^{\rm L})$ relation. Since the values assumed by the $b_{\nabla^2\delta}^{\rm L}(b_1^{\rm L})$ relation depend on our choice of $k_{\rm max}$ and smoothing scale $k_{\rm d}$, the fitting function for $b_{\nabla^2\delta}^{\rm L}(b_1^{\rm L})$ shown in Table \ref{tab:polyfits} and the allowed ranged reported in Eq. (\ref{eq::priors}) can only be used with $k_{\rm max} = 0.7~\ihMpc$ and $k_{\rm d} = 0.75 ~ \ihMpc$. Any other fitting configuration should require leaving $b_{\nabla^2\delta}^{\rm L}$ free.

\begin{table}
    \centering
    \begin{tabular}{rl}
        \hline
        \multicolumn{2}{c}{Haloes, $k_{\rm d} = 0.75 \ihMpc$, $k_{\rm max} = 0.7 \ihMpc$}\\
        \hline
        $b_{2}^{\rm L}(b_1^{\rm L})=$ & $-0.09143 (b_1^{\rm L})^3 + 0.7093 (b_1^{\rm L})^2 - 0.2607 b_1^{\rm L} - 0.3469$\\
        \hline
        $b_{s^2}^{\rm L}(b_1^{\rm L})=$ & $0.02278 (b_1^{\rm L})^3 - 0.005503 (b_1^{\rm L})^2 - 0.5904 b_1^{\rm L} - 0.1174$\\
        \hline
        $b_{\nabla^2\delta}^{\rm L}(b_1^{\rm L})=$ & $-0.6971 (b_1^{\rm L})^3 + 0.7892 (b_1^{\rm L})^2 + 0.5882 b_1^{\rm L} - 0.1072$\\
        \hline
        \hline
        \multicolumn{2}{c}{Galaxies, $k_{\rm d} = 0.75 \ihMpc$, $k_{\rm max} = 0.7 \ihMpc$}\\
        \hline
        $b_{2}^{\rm L}(b_1^{\rm L})=$ & $0.01677 (b_1^{\rm L})^3 - 0.005116 (b_1^{\rm L})^2 + 0.4279 b_1^{\rm L} - 0.1635$\\
        \hline
        $b_{s^2}^{\rm L}(b_1^{\rm L})=$ & $-0.3605 (b_1^{\rm L})^3 + 0.5649 (b_1^{\rm L})^2 - 0.1412 b_1^{\rm L} - 0.01318$\\
        \hline
        $b_{\nabla^2\delta}^{\rm L}(b_1^{\rm L})=$ & $0.2298 (b_1^{\rm L})^3 - 2.096 (b_1^{\rm L})^2 + 0.7816 b_1^{\rm L} - 0.1545$\\
        \hline
    \end{tabular}
    \caption{The polynomial fitting functions obtained for haloes and galaxies from our simulations in the context of the hybrid Lagrangian bias expansion model. We consider our fiducial choice of smoothing $k_{\rm d} = 0.75 \, \ihMpc$ and $k_{\rm max} = 0.7 \, \ihMpc$.}
    \label{tab:polyfits}
\end{table}

\section{Discussion}\label{sec::discussion}
In the last part of this work, we focus on exploring potential causes for the difference between the bias relations of galaxies and the ones of halos. We identify two effects, both linked to how galaxies occupy haloes: on the one hand, we consider that, while these relations for haloes have been obtained splitting haloes in differential mass bins, our galaxy samples correspond to cumulative bins; these reflect the averaged behavior of the bias of different host mass bins and are influenced by how different haloes host different galaxy populations according to different halo occupation distributions (HODs). On the other hand, even for the same HOD and the same cumulative behaviour, we consider the effect of galaxy assembly bias (GAB), i.e. the occupancy variation induced by the dependence of the HOD on properties other than the host halo mass.

To assess the effect of different HODs on the resulting average bias, we employ the concept of effective bias \citep{BensonEtal2000}. Supposing we know the halo mass function $n_{\rm h}(M_{\rm h})$, the number of galaxies in each halo of a given mass, i.e the HOD $n_{\rm g}(M_{\rm h})$, and the bias-mass relation in differential mass bins $b_i(M_{\rm h})$, then the effective value of a given order bias parameter is
\begin{equation}
    \tilde{b_i} = \dfrac{\int \dif M_h n_{\rm h}(M_{\rm h}) n_{\rm g}(M_{\rm h}) b_i(M_{\rm h})}{\int \dif M_h n_{\rm h}(M_{\rm h}) n_{\rm g}(M_{\rm h})},
    \label{eq::effbias}
\end{equation}
with the integration range covering the span of available host halo masses in the sample.

We assume the mass function to be described by the fitting formula of \cite{Ondaro-MalleaEtal2021}. We simplify our problem assuming a standard 5-parameter model for the HOD (which is known to be a good description of SM selected galaxies but does not reproduce the HOD of SFR selected galaxies). The total number of galaxies in haloes of mass $M_{\rm h }$ is obtained by separately modelling the number of centrals
\begin{equation}
    N_{\rm cen} = \dfrac{1}{2} \left[1 + \mathrm{erf}\left(\dfrac{\log M_{\rm h} - \log M_{\rm min}}{\sigma_{\log M}}\right)\right],
\end{equation}
and of satellite galaxies
\begin{equation}
    N_{\rm sat} = N_{\rm cen} \left[\dfrac{M_{\rm h} - M_0}{M_1}\right]^\alpha,
\end{equation}
being $M_{\rm min}, \sigma_{\log_M}, M_{1}, M_{0}$ and $\alpha$ the free parameters of the model.

\begin{figure}
    \includegraphics[width=0.48\textwidth]{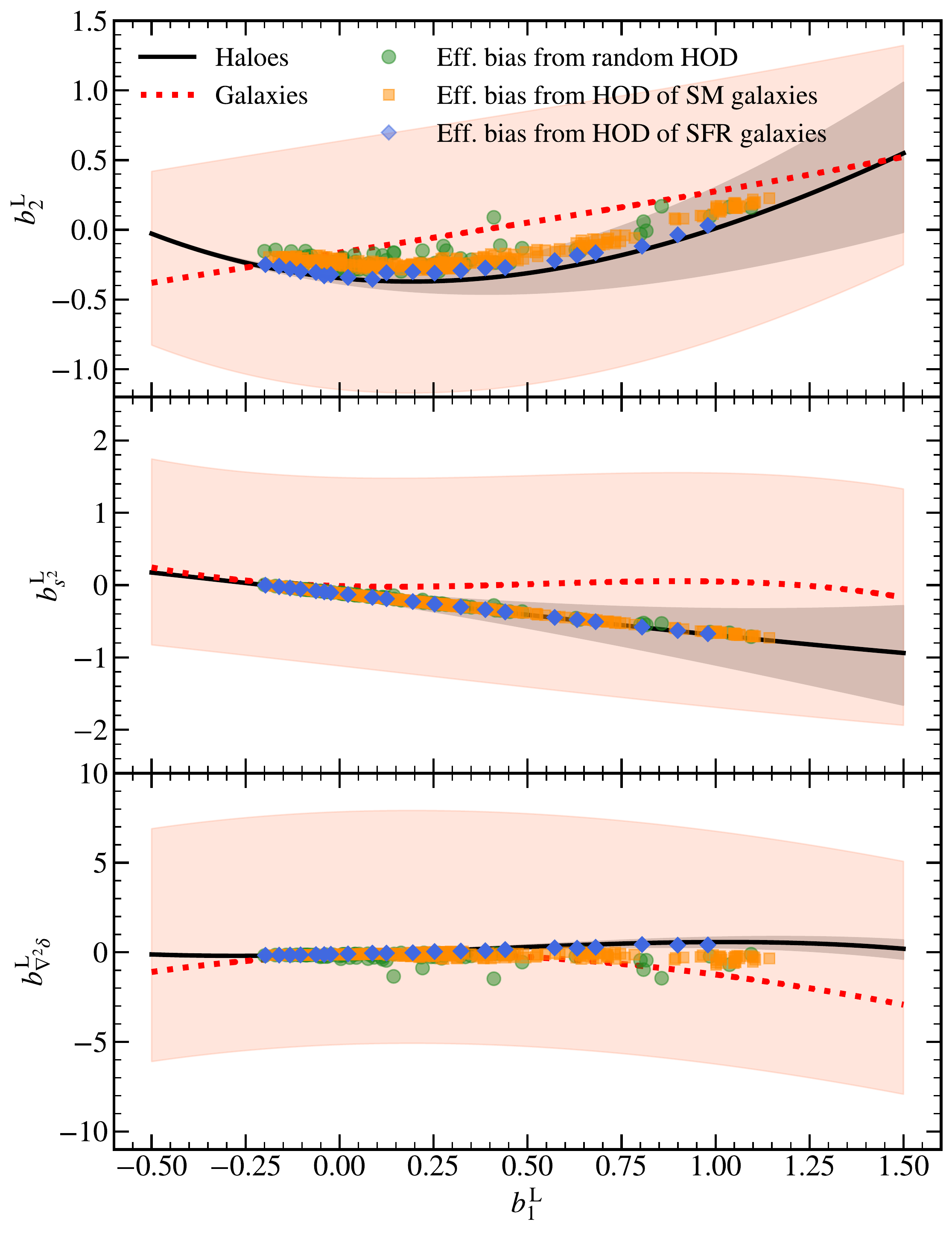}
    \caption{The dependence of the $b^{\rm L}_i(b^{\rm L}_1)$ relation on the sample HOD. Each point represents the effective bias corresponding to a given choice of HOD parameters, according to Eq. (\ref{eq::effbias}). In the case of green circles, the HOD parameters are chosen randomly. For orange squares, instead, the HOD parameters have been fitted to the SM selected galaxy samples of our smaller size simulations (at $z=0$ and $1$). For blue diamonds instead we use the HOD measured from SFR selected galaxies. The black and red lines are the polynomial fits to our halo and galaxy samples (respectively) as presented in Tab. \ref{tab:polyfits}. We also show the galaxy priors from Eq. (\ref{eq::priors}) as a red shaded area.}
    \label{fig::effbias}
\end{figure}

We create a random sampling of the HOD free parameters with a 5D latin hypercube width sides
\begin{displaymath}
    \begin{split}
        \log_{10} M_{\rm min} &\in [11, 13],\\
        \sigma_{\log_M} &\in [0, 0.9],\\
        M_{1} / M_{\rm min} &\in [1, 30],\\
        \log_{10} M_{0} &\in [11.5, 14],\\
        \alpha &\in [0.8, 1.2].\\
    \end{split}
\end{displaymath}
In Fig. \ref{fig::effbias} we show the effect of changing the HOD parameters (for exactly the same mass function) assuming these random HOD parameters. The $b_2^{\rm L}(b_1^{\rm L})$ relation moves away from the prediction for haloes, and scattering appears, even if we cannot reproduce entirely the relation obtained for galaxies. Interestingly, the relations $b_{s^2}^{\rm L}(b_1^{\rm L})$ and $b_{\nabla^2\delta}^{\rm L}(b_1^{\rm L})$ obtained considering the effective bias are extremely tight around the fits obtained from haloes. This is probably due to the fact that the $b^{\rm L}_i(\nu)$ relations are almost constant for these parameters (see Fig. \ref{fig::halo-mass}) and effects other than the HOD must be responsible for the scatter we see for galaxies.

Moreover, we fit the HOD parameters to SM-selected galaxy samples using the smaller size simulations presented in section \ref{sec::sims-mocks} to make the analysis more computationally manageable. We also use the measured (in this case, not fitted) HOD of SFR selected galaxies from these same simulations. We compute the effective bias using these fitted HODs (orange and blue points in figure \ref{fig::effbias}). In this case, the level of scatter is significantly reduced, especially for the SM selected sample.

Finally, we consider the effect of GAB. We expect our galaxy catalogues to include GAB, since we created them using a SHAMe technique. There are two ways to single out the effect of GAB: on the one hand, one can shuffle all galaxies of a sample in narrow bins of host halo mass, thus washing out any dependency of galaxy occupancy other than halo mass; on the other end, one can fit the HOD parameters to the galaxy sample and then use those parameters to create a new galaxy catalogue that by construction is agnostic of dependencies other than host halo mass. While we tested both methods with identical results, we present here the latter.

\begin{figure}
    \includegraphics[width=0.48\textwidth]{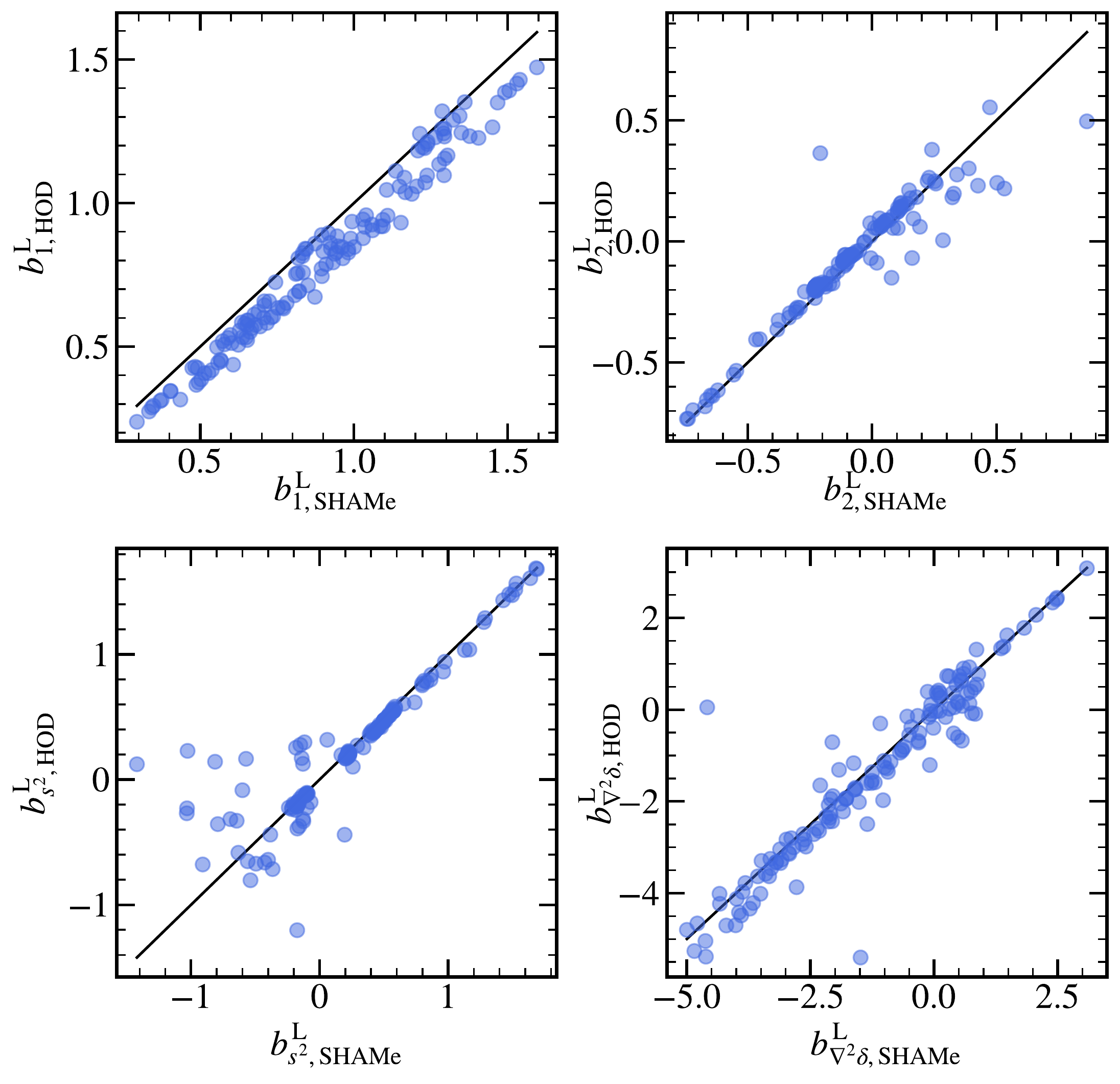}
    \caption{The effect of galaxy assembly bias on the Lagrangian bias relations for SM selected galaxies from the smaller-size simulations. In the x-axis the bias parameters come from fitting the clustering of galaxy samples created with the SHAMe technique, which therefore include a variable amount of GAB. The y-axis, instead, refers to the bias parameters obtained fitting mock catalogues built using HOD parameters fitted to the SM selected galaxies (so this samples of galaxies have the same HODs as the SHAMe ones, but no GAB).}
    \label{fig::gab}
\end{figure}

In particular, in Fig. \ref{fig::gab} we show the bias relations obtained for the SM-selected galaxy samples using, once again, the smaller size simulations. At the same time, we also show the bias relations obtained for a sample of HOD galaxies, whose HOD parameters have been obtained from the SHAMe galaxies. We can see that, once removed the effect of GAB, the linear bias considerably shifts towards smaller values. On the contrary, the changes in $b_2^{\rm L}$, $b_{s^2}^{\rm L}$ and $b_{\nabla^2\delta}^{\rm L}$ are not particularly significant. We interpret this by considering that GAB by definition affects the large scale clustering of galaxies, and is therefore expected to appear mostly in the parameter $b_1^{\rm L}$.

While for dark matter haloes the effect of halo assembly bias on higher order bias parameters (in Eulerian bias expansions) is well established \citep{AnguloBaughLacey2008,LazeyrasMussoSchmidt2017}, there is still no consensus on the effect of GAB. However, our findings for galaxies qualitatively agree with those of \cite{LazeyrasBarreiraSchmidt2021}, where the authors, focussing on dark matter haloes, find almost no impact of AB on the $b_2(b_1)$ relation. They do, however, find a significant effect of halo assembly bias on the $b_{s^2}(b_1)$ relation. On our part, we notice an increased scatter among our points for $b_{s^2}^{\rm L}$ in Fig. \ref{fig::gab}, but, given the uncertainties of our fitting method, we cannot draw any definitive conclusion on this matter.

As a final remark, we conclude that both variations in the HOD and in GAB can partly explain the difference between the coevolution relations of galaxies and haloes, both in terms of systematic shifts and scatter. However, other effects must be in place to fully explain this difference. For instance, \cite{VoivodicBarreira2020} proposed an extension of the halo model that considers the response of the HOD to long-wavelength perturbations. In \cite{BarreiraLazeyrasSchmidt2021}, the authors have shown that these HOD responses can also induce differences between the coevolution relations of haloes and those of galaxies, and should therefore be taken into account.

\section{Conclusions}\label{sec::conclu}
In this paper we have assumed a hybrid second order Lagrangian bias expansion and inferred the bias parameter for many different bias tracers from $N$-body simulations: we have considered dark matter haloes, SM-selected galaxies and SFR-selected galaxies. We have obtained the galaxy samples employing an extended SHAM technique.
\begin{itemize}
    \item We find that the hybrid Lagrangian bias model is a good description of generic bias tracers. We show in Figs. \ref{fig::bestfit-model-contours}, \ref{fig::bestfit-kmax} and \ref{fig::bestfit-model} that it can fit our fiducial SFR-selected sample, in a way that is robust against different choices of smoothing scale in Lagrangian space and maximum wavenumber included in the fit. Moreover, in Fig. \ref{fig::chi2s} we show that this model can describe thousands of different galaxy populations with reduced $\chi^2 < 1$.
    \item By analysing haloes in differential mass bins, we find that the values of the different bias parameters inferred with our method are roughly compatible with commonly used fitting functions. However, we notice some small systematic discrepancies in the $b_2^{\rm L}(\nu)$ relation for $\nu > 1$. These departures from theoretical predictions and fitting formulae are likely due to approximations in the connection between Eulerian and Lagrangian parameters, but could in principle also come from different assumptions in the modelling of biased tracers and scales included in the fits.
    \item The systematic differences between our bias parameters and other theoretical predictions for haloes are present also when considering the coevolution relations $b_i^{\rm L}(b_1^{\rm L})$. We find that the coevolution relations obtained for our biasing model are robust against changes in $k_{\rm max}$ and smoothing for the linear and quadratic biases, while depend on the details of the fit for the higher-derivative bias (Fig. \ref{fig::kmax-halos}).
    \item We find that galaxy bias parameters follow $b_i^{\rm L}(b_1^{\rm L})$ relations that are different from those of haloes in differential mass bins (assuming the same biasing model). In particular, compared to haloes, galaxies show systematically higher $b_2^{\rm L}$ at fixed $b_1^{\rm L}$. Moreover the coevolution relations of $b_{s^2}^{\rm L}$ and $b_{\nabla^2\delta}^{\rm L}$ exhibit larger scatter than their halo counterparts. In Figs. \ref{fig::b1-b2-sm} and \ref{fig::b1-b2-sfr} we study these differences singling out the effects of the galaxy formation parameters, satellite fraction and cosmology. We find that the bias relations depend non trivially on the galaxy formation model assumed. In Fig. \ref{fig::b1-b2-summary} we present all of our models together, to have a grasp of the bias parameter space spanned by our realistic galaxy samples.
    \item We link the shifts and scatter in the bias parameter relations between the case of haloes in mass bins and the cumulative galaxy bins to at least two causes: on the one hand, using ad-hoc HOD mocks we show that changing the way galaxies populate haloes does introduce a scatter in the $b_2^{\rm L}(b_1^{\rm L})$ relation and systematically increases $b_2^{\rm L}$ (Fig. \ref{fig::effbias}). However, the differences in HOD alone seem not to be enough to explain the difference between coevolution relations between galaxies and haloes (especially the scattering for the $b_{s^2}^{\rm L}$ and $b_{\nabla^2\delta}^{\rm L}$ relations). On the other hand, by erasing the effect of assembly bias from our galaxy samples we show that galaxy assembly affects almost exclusively $b_1^{\rm L}$. We conclude that the difference between the coevolution relations for haloes and galaxies in the context of the hybrid Lagrangian bias expansion model are partly due to the different halo occupations of different galaxy samples and to occupancy variations for samples with same HOD. However, other processes might also be important such as those that set the spatial distribution of satellite galaxies inside haloes.
\end{itemize}
These results illustrate the typical values assumed by Lagrangian bias parameters corresponding to very different galaxy populations and number densities, between $z=0$ and $1$.  We provide in Table \ref{tab:polyfits} the fitting functions for the coevolution relations for bias parameters in the context of the hybrid Lagrangian bias expansion model (at second order) for both haloes and galaxies. Moreover, in Eq. (\ref{eq::priors}) we provide the boundaries of a hypervolume enclosing the values of bias parameters describing all of our galaxy samples. We anticipate that these formulae can be used as a prior for future bayesian analyses where both bias parameters and cosmological parameters are constrained, thus reducing the size of the parameter space to be explored. In the future we expect that an even better characterisation of these coevolution relations will be possible combining the power spectra analysis with other observables and, in particular, including higher order correlations. Moreover, we expect that including third order terms in the bias expansion will be key to determine the actual range of applicability of these relations to fitting configurations different from the ones assumed in this work.

\section*{Acknowledgements}
The authors acknowledge the support of the ERC-StG number 716151 (BACCO). MPI acknowledges the support of the ``Juan de la Cierva Formaci\'on'' fellowship (FJC2019-040814-I). FGM acknowledges support from FAPESP via the fellowship 2019/01631-0. The authors acknowledge the computer resources at MareNostrum and the technical support provided by Barcelona Supercomputing Center (RES-AECT-2019-2-0012, RES-AECT-2020-3-0014).

\section*{Data Availability}
The data underlying this article will be shared on reasonable request to the corresponding author. The galaxy power spectra will be available at \url{https://bacco.dipc.org/galpk.html}.



\bibliographystyle{mnras}
\bibliography{Bibliography_all} 







\bsp	
\label{lastpage}
\end{document}